\documentclass[structabstract]{aa}  
\usepackage{graphicx}
\usepackage{lscape}
\usepackage{amsmath}
\usepackage{txfonts}
\usepackage{url}
\usepackage{natbib}
\usepackage{xcolor}
\bibpunct{(}{)}{;}{a}{}{,}

\begin{document}
\title{Populations of rotating stars}

\subtitle{I. Models from $1.7$ to $15\,M_{\sun}$ at $Z = 0.014$, $0.006$, and $0.002$\\with $\Omega/\Omega_\text{crit}$ between $0$ and $1$}

\author{C. Georgy\inst{1,2}, S. Ekstr\"om\inst{3}, A. Granada\inst{3}, G. Meynet\inst{3}, N. Mowlavi\inst{3}, P. Eggenberger\inst{3}, \and A. Maeder\inst{3}}

\authorrunning{Georgy et al.}

\institute{Astrophysics group, EPSAM, Keele University, Lennard-Jones Labs, Keele, ST5 5BG, UK\\
                   \email{c.georgy@keele.ac.uk} \and
                   Centre de Recherche Astrophysique de Lyon, Ecole Normale Sup\'erieure de Lyon, 46, all\'ee d'Italie, F-69384 Lyon cedex 07, France
                 \and Geneva Observatory, University of Geneva, Maillettes 51, CH-1290 Sauverny, Switzerland}

\date{Received ; accepted }

 
\abstract
{B-type stars are known to rotate at various velocities, including very fast rotators near the critical velocity as the Be stars.} 
{In this paper, we provide  stellar models covering the mass range between $1.7$ to $15\,M_{\sun}$, which includes the typical mass of known Be stars, at $Z=0.014$, $0.006$, and $0.002$ and for an extended range of initial velocities on the zero-age main sequence.}
{We used the Geneva stellar-evolution code, including the effects of shellular rotation, with a numerical treatment that has been improved so the code can precisely track the variation in the angular momentum content of the star as it changes under the influence of radiative winds and/or mechanical mass loss.}
{We discuss the impact of the initial rotation rate on the tracks in the Hertzsprung-Russell diagram, the main-sequence (MS) lifetimes, the evolution of the surface rotation and abundances, as well as on the ejected masses of various isotopes. Among the new results obtained from the present grid we find that 1) fast-rotating stars with initial masses around $1.7\,M_{\sun}$ present at the beginning of the core hydrogen-burning phase quite small convective cores with respect to their slowly rotating counterparts. This fact may be interesting to keep in mind in the framework of the asteroseismic studies of such stars. 2) The contrast between the core and surface angular velocity is higher in slower rotating stars. The values presently obtained are in agreement with the very few values obtained for B-type stars from asteroseismology. 3) At $Z=0.002$, the stars in the mass range of $1.7$ to $3\, M_{\sun}$ with a mean velocity on the MS of the order of $150\,\text{km}\cdot\text{s}^{-1}$ show N/H enhancement superior to $0.2\,\text{dex}$ at mid-MS, and superior to $0.4\,\text{dex}$ at the end of the MS phase. At solar metallicity the corresponding values are below $0.2\,\text{dex}$ at any time in the MS.}
{An extended database of stellar models containing 270 evolutionary tracks is provided to the community.}


\maketitle

\section{Introduction}

In the framework of the new grids of stellar models that the Geneva stellar evolution group has recently made available to the community \citep[][hereafter Grids I]{Ekstrom2012a}\defcitealias{Ekstrom2012a}{Grids I}\!, we propose a more detailed study of some subsamples of the full grids, offering a reduced mass range, but a more extended coverage in initial rotation rates.

An interesting subsample is the mass domain between $1.7$ to $15\,M_{\sun}$, which corresponds roughly to the domain of the early A- and B-type stars. This range of mass covers the transition from the massive and energetic O-stars, dominated by their stellar winds, to the less massive A stars with negligible winds. Then, in the B-type star range, the interplay between stellar winds and rotation will together affect the evolution of the star and the injection of mass and energy to the circumstellar environment, and also determine the post-main-sequence fate of the star.  Moreover, this range contains the interesting case of Be-type stars. Since rapid rotation and reaching the critical velocity is thought at least partially to be involved to explain the Be-star phenomenon \citep{Martayan2007a,Martayan2010a}, we provide in this grid an extended coverage of initial rotations up to the fastest cases.

Previous studies have already explored the evolution of B stars towards the critical velocity as a function of the metallicity \citep{Ekstrom2008b}. However, the Geneva stellar-evolution code at that time could not follow the model through a critical-rotation phase while keeping precise track of the stellar angular-momentum content, and computation of the models was stopped when they reached the beginning of that stage. Several recent improvements in the Geneva code now make it possible to compute models of critically rotating stars, and to give a theoretical estimation of the equatorial mass loss that such stars undergo during this phase. In that context, we performed a new analysis of stars in the range of $1.7$ to $15\,M_{\sun}$ at various metallicities: $Z=0.014$ (solar), $Z=0.006$ (LMC), and $Z=0.002$ (SMC), with initial rotation rates on the zero-age main sequence (ZAMS)  between $0<\Omega/\Omega_\text{crit}<0.95$.

The present theoretical database will allow synthetic clusters to be built with the inclusion of any distribution of initial rotational velocities. In a forthcoming paper, we shall discuss our population-synthesis code and compare the results with observed stellar clusters and field star populations.

This paper is organised as follows. In Sect.~\ref{Numerics}, we briefly recall the effect of the rotation on the stellar surface properties. We explain the numerical method used to estimate the equatorial mass loss for critically-rotating stars. In Sect.~\ref{Physics}, we briefly describe the parameters and the physical ingredients used to perform our numerical simulations. The results are presented in Sect.~\ref{Results}. Some aspects of the advanced phases are discussed in Sect.~\ref{Sadvanced}. Finally, Sect.~\ref{Conclu} contains our conclusions.

A complete discussion concerning our faster rotating models and comparison with observations of Be-type stars is the subject of the second paper of this series \citep[][hereafter Paper II]{Granada2012a}.
\defcitealias{Granada2012a}{Paper II}

\section{Angular momentum losses due to stellar winds and equatorial mechanical mass losses: numerical method}\label{Numerics}
\subsection{Recall of the effects of rotation on the properties of the stellar surface}

Rotation strongly affects the stellar surface in several ways. First of all, the shape of the surface itself is modified: the equatorial radius becoming larger than the polar one under the action of the centrifugal force. The shape of the star depends on several parameters \citep[ratio to the critical velocity, latitudinal differential rotation, etc., see][]{Zorec2011a}. Here, given the shellular hypothesis used in our stellar-evolution code \citepalias[see][for details]{Ekstrom2012a} the shape of the star is accounted for in the frame of the Roche model, as it allows for a quick numerical treatment. In this context, the only pertinent parameter for describing the stellar surface is the ratio of the actual angular velocity to the critical one $\omega = \Omega/\Omega_\text{crit}$ \citep[see \textit{e.g.}][]{Georgy2011a}. The parameter $\Omega_\text{crit}$ is defined as
\begin{equation}
\Omega_\text{crit} = \sqrt{\frac{GM}{R^3_\text{e, crit}}},
\label{EqOmegaCrit}
\end{equation}
with $R_\text{e, crit}$ the equatorial radius when the star rotates at the angular velocity $\Omega_\text{crit}$. In the framework of this model, the maximum ratio of the equatorial-to-polar radius (when $\Omega = \Omega_\text{crit}$) is $1.5$. This is observationally supported by the observations of the fast-rotating star Achernar \citep{Vinicius2006a,Carciofi2008a}.

According to \citet{Maeder2000a}, the total acceleration at the surface of a rotating star, including gravity, radiation-pressure acceleration and centrifugal acceleration can be expressed as (vectors are in boldface)
\begin{equation}
\mathbf{g}_\text{tot}  = \mathbf{g}_\text{eff}\left(1-\Gamma(\Omega,\theta)\right),\label{Eqgtot}
\end{equation}
with $\mathbf{g}_\text{eff} = \mathbf{g}_\text{grav} + \mathbf{g}_\text{cen}$ the sum of the gravitational and centrifugal accelerations, and $\Gamma(\Omega,\theta)$ the local Eddington factor, accounting for the effects of the rotation. Following \citet{Maeder2000a}, we define a velocity that vanishes the expression (\ref{Eqgtot}) as a critical velocity. The \textit{first critical velocity} is given by $\mathbf{g}_\text{eff} = 0$, and can be expressed as
\begin{equation}
v_\text{crit, 1} = \sqrt{\frac{2}{3}\frac{GM}{R_\text{p, crit}}},\label{EqVcrit}
\end{equation}
with $R_\text{p, crit}$ being the polar radius when the star reaches $\Omega_\text{crit}$. The \textit{second critical velocity} is reached when $1-\Gamma(\Omega,\theta) = 0$. It is only possible for stars with a very high Eddington factor ($\Gamma_\text{Edd} \gtrsim 0.639$). In this work, all the considered stellar models remain far from this value, and the only pertinent critical velocity to consider is the first one. In the following text, we systematically use the term ``critical velocity'' for ``first critical velocity''.

Rotation not only affects the shape of the surface, but also produces latitudinal variations in the radiative flux, effective temperature, and mass flux, producing an anisotropic stellar wind \citep[see \textit{e.g.}][]{Maeder2000a,Maeder2002a}. The effects of such a wind on the evolution of stars remaining far from the critical velocity (which is the case for most of stars) are moderate \citep{Georgy2011a}, and can usually be neglected. In this paper, however, some of our models evolve very close to or even at the critical velocity. The effects of anisotropic winds are thus stronger \citep[more than $10\%$ in terms of angular-momentum loss compared to models where such effects are neglected, see][]{Georgy2011a}, and are accounted for in this work. Also, wind anisotropies may produce, at least for some periods, non-spherical  shape bubbles around rapidly-rotating stars, a feature that may be interesting for comparisons with observations.

\subsection{Computation of the equatorial mechanical mass loss}\label{SSMechanicalMassLoss}

During the stellar lifetime, both the surface angular velocity and the critical one evolve, making the rotation rate $\omega=\Omega_\text{surf}/\Omega_\text{crit}$ change as a function of time. It is thus possible for a star to reach the critical velocity during its life, even if it started on the ZAMS with $\omega_\text{ini}<1$. Once this limit is attained, the effective gravity at the equator of the star is nil. We thus expect a strong enhancement of the mass loss in the equatorial region, hereafter called ``mechanical mass loss'', which removes the overcritical layers and maintains the surface at the critical velocity or slightly below. The exact details of this process is to date not well known, and would need complex simulations that combine hydrodynamics and radiative transfer. It is currently not possible to perform such computations coupled with a stellar-evolution code for the full duration of the stellar life.

To account for this mechanical mass loss in our stellar-evolution code, we proceed as follows\footnote{This approach is similar to the one described in \citetalias{Ekstrom2012a} ensuring the conservation of angular momentum.}: we estimate the angular-momentum amount $\Delta\mathcal{L}_\text{rad}$ that the model loses through radiative stellar winds during a time step, on the basis of the estimated surface quantities (radius, mass-loss rate, \textit{etc.}). If the effects of the anisotropic stellar winds are not accounted for, we have
\begin{equation}
\Delta\mathcal{L}_\text{rad} = \frac{2}{3}\Delta M_\text{rad}\Omega_\text{surf} r_\text{surf}^2,
\end{equation}
with $\Delta M_\text{rad}$ the total amount of mass lost during the time step due to radiative stellar winds, and $r_\text{surf}$ the radius of the star. (This mean radius is defined as the radius that would have a spherical star having the same luminosity and effective temperature.) If the wind anisotropy is accounted for, this expression is more complex and needs a numerical integration over the stellar surface \citep[see][]{Georgy2011a}.

Knowing the initial characteristics of the model at the beginning of the time step, we have to estimate whether the radiative mass loss is sufficient to keep the star below the critical velocity at the end of the time step. Moreover, because of numerical difficulties, it is not possible to maintain the model exactly at the critical velocity. We thus define a maximal ratio $\omega_\text{max} \equiv \frac{\Omega_\text{max allowed}}{\Omega_\text{crit}}$. In our computations, it is set to $\omega_\text{max} = 0.99$. 

At any time, the structure of our numerical model is composed of an envelope, where convection is non-adiabatic and ionisation not complete, and of an interior zone where convection is adiabatic and ionisation complete. The envelope, which encompasses a very small fraction of the total mass, typically one ten-thousandth, is assumed to rotate with the same angular velocity as the uppermost layers of the interior zone\footnote{\footnotesize{Note that during the red (super)giant phase, the envelope is much bigger than during the MS. When the star crosses the Hertzsprung-Russell diagram and becomes red, the mass-coordinate of the base of the envelope is progressively decreased, up to $M_r/M_\text{tot} = 0.98$.}}. The mass removed by the mass loss will be a fraction of the mass of the envelope. The removal of these layers will remove angular momentum and a new angular velocity distribution inside the star will be built up so that the total angular momentum of the star will be decreased by the exact amount lost by the star. Since angular momentum can be transported over a given distance in a time step, the zone where $\Omega$ is modified covers a zone with a depth approximately equal to the radial component of the meridional current multiplied by the time step. We thus have to compute the correcting factors of $\Omega$ in that zone so that the process keeps precise track of the angular momentum. In the following, we denote $N_\text{corr}$ as the number of layers that are affected by the mass loss. To estimate the correcting factors, called $q$ below, we suppose that:
\begin{itemize}
\item the changes in the structure of the star are small enough during the time step considered that they can be neglected for computing the $q$ values;
\item the angular momentum removed by the radiative stellar wind is equally removed from the envelope and the first $N_\text{corr}$ layers (see \citealt{Georgy2010a} and \citetalias{Ekstrom2012a} for more details). That means that $q$ is constant over the whole zone where $\Omega$ will be affected by the mass loss, or that $\Omega_\text{i, new} = \Omega_\text{i, ini} \left(1 +q\right)$, with $i$ the number of the shell considered;
\item no other angular momentum transport mechanisms act during the time step.
\end{itemize}
The initial angular momentum content of the $N_\text{corr}$ layers is $\mathcal{L}_\text{tot, ini} = \mathcal{L}_\text{e, ini} + \sum_{i=1}^{N_\text{corr}}\mathcal{L}_{i\text{, ini}}$, where the subscript ``e'' is for the envelope, and $i$ concerns the \textit{i}-th layers below the surface. Assuming that the mass carried away by the stellar winds is removed in the envelope, we can thus write the estimated total amount of angular momentum in the $N_\text{corr}$  layers after the mass loss occurred as
\begin{equation}
\mathcal{L}_\text{tot, fin} = \mathcal{L}_\text{e, ini}\left(1 + q\right)\left(1-\frac{\Delta M_\text{rad}}{m_\text{e, ini}}\right) + \left(1+q\right)\sum_{i=1}^{N_\text{corr}}\mathcal{L}_{i\text{, ini}},
\end{equation}
with $m_\text{e, ini}$ the mass of the envelope at the beginning of the time step. Finally, since we must have $\mathcal{L}_\text{tot, fin} = \mathcal{L}_\text{tot, ini} - \Delta\mathcal{L}_\text{rad}$, we obtain
\begin{equation}
q = \frac{\mathcal{L}_\text{e, ini}\frac{\Delta M_\text{rad}}{m_\text{e, ini}}-\Delta\mathcal{L}_\text{rad}}{\mathcal{L}_\text{tot, ini} - \mathcal{L}_\text{e, ini}\frac{\Delta M_\text{rad}}{m_\text{e, ini}}}.
\end{equation}
Knowing $q$, it is now possible to estimate the surface angular velocity $\Omega_\text{surf, new}$ at the end of the time step. If $\Omega_\text{surf, new} < \Omega_\text{max allowed}$, the model does not need to lose more angular momentum and the angular momentum losses estimate is complete.

If $\Omega_\text{surf, new} > \Omega_\text{max allowed}$, the radiative mass loss is not strong enough to maintain the star below the critical velocity, and other expressions for $\mathcal{L}_\text{tot, fin}$ and $q$ have to be obtained. In contrast to previous case, where the mass loss rate imposes an angular momentum loss, we need here to remove an excess of angular momentum, and this will determine an additional mass loss.

We assume that this additional mass, $\Delta M_\text{mech}$, is lost mechanically at the equator, and rotates at the same angular velocity as the surface. It thus removes the following angular momentum:
\begin{equation}
\Delta\mathcal{L}_\text{mech} = \Delta M_\text{mech}\Omega_\text{surf} r_\text{eq}^2,\label{EqDLmec}
\end{equation}
with $ r_\text{eq}$ the equatorial radius. The mechanical mass loss has to bring the angular velocity of the stellar surface down to $\Omega_\text{max allowed}$. We thus define the factor $q_\text{lim}$ as $\Omega_\text{surf, new} = \Omega_\text{max allowed} = \Omega_\text{surf, ini} \left(1 +q_\text{lim}\right)$, and the new angular velocity of each layer where the correction is applied becomes
\begin{equation}
\Omega_{i\text{, new}} = \Omega_{i\text{, ini}} \left(1 +q_\text{lim}\right) = \Omega_{i\text{, ini}} \frac{\Omega_\text{max allowed}}{\Omega_\text{surf, ini}}.\label{EqOmega}
\end{equation}

As before, we assume that the effect of the mechanical mass loss will affect $N_\text{corr}$ layers below the surface. The angular momentum of the star that is needed to have a sub-critically rotating surface at the end of the time step is estimated to be
\begin{align}
\mathcal{L}_\text{tot, lim} &=  \mathcal{L}_\text{e, ini}\left(1 + q_\text{lim}\right)\left(1-\frac{\Delta M_\text{rad}+\Delta M_\text{mech}}{m_\text{e, ini}}\right) \notag\\
&+ \left(1+q_\text{lim}\right)\sum_{i=1}^{N_\text{corr}} \mathcal{L}_{i\text{, ini}}.
\end{align}
Knowing that $\mathcal{L}_\text{tot, lim} = \mathcal{L}_\text{tot, ini} - \Delta\mathcal{L}_\text{rad} - \Delta\mathcal{L}_\text{mech}$, and using relations (\ref{EqDLmec}) and (\ref{EqOmega}), the mechanical mass lost during the time step $\Delta M_\text{mech}$, which is the only unknown of the equations, can now be defined as
\begin{equation}
\Delta M_\text{mech} = \frac{\mathcal{L}_\text{tot, ini}\left(1-\frac{\Omega_\text{max}}{\Omega_\text{surf, ini}}\right) + \mathcal{L}_\text{e, ini}\frac{\Delta M_\text{rad}}{m_\text{e, ini}}\frac{\Omega_\text{max}}{\Omega_\text{surf, ini}} - \Delta\mathcal{L}_\text{rad}}{\Omega_\text{surf, ini} r_\text{eq}^2 - \frac{\mathcal{L}_\text{e, ini}}{m_\text{e, ini}}\frac{\Omega_\text{max}}{\Omega_\text{surf, ini}}},
\end{equation}
providing an estimation on how much mass should be removed from the equator of the star in order to keep the star below the critical velocity.  This mass will probably form an equatorial circumstellar disc. In this paper, we assume that this mass is lost, and is not re-accreted on the star.

As in \citetalias{Ekstrom2012a}, $N_\text{corr}$ is set to 200 in this work. This value ensures that during a characteristic time step, the diffusion and the advection of the angular momentum affect at least these layers. To avoid a strong discontinuity at the edge of the zone where the correction is applied, we limited the time step in order to limit the value of $q$ to $0.005$ ($0.01$ for models rotating with $\omega < 0.2$). Moreover, during the red (super)giant phase, during which a large surface convective zone develops, the correction is applied only to the whole convective zone, as soon as it stretches over more than 50 layers.

The above method for computing the variation in $\Omega$ due to the mass loss, while allowing the conservation of the angular momentum, is probably quite schematic, especially for what concerns the equatorial mass loss when the surface velocity becomes overcritical. Among the limitations, we can mention that the process that ultimately pushes the matter out is absent. This process can be momentum input by radiation and/or pulsations and/or convection. To fully describe such processes, other tools than the present stellar-evolution code are necessary. The discussion about to what extent the results of the present models depend on the computational time step and on the value of $\omega_\text{max}$ is presented in \citetalias{Granada2012a}.

\section{Physics of the models}\label{Physics}

\subsection{General context}

\begin{figure*}
\begin{center}
\includegraphics[width=\textwidth]{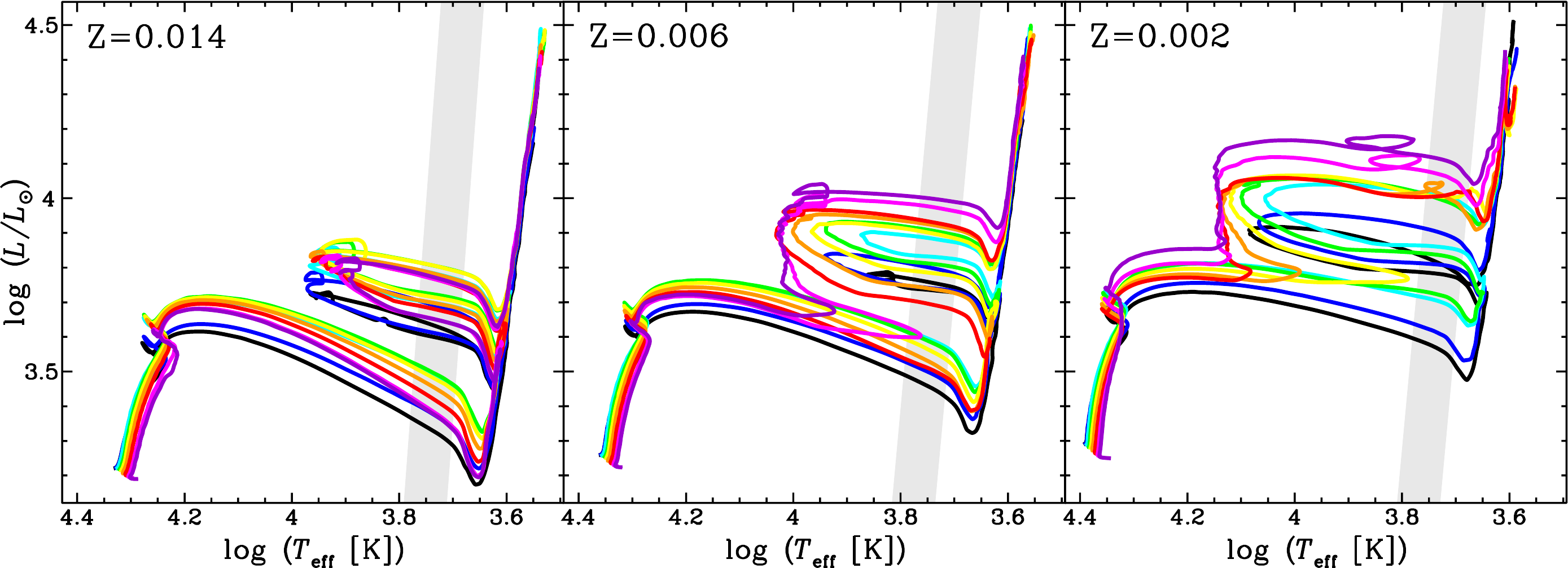}
\end{center}
\caption{HRD for the 7 $M_{\sun}$ models with $\omega_\text{ini}=\Omega_\text{ini}/\Omega_\text{crit}=$ 0 (black), 0.1 (blue), 0.3 (cyan), 0.5 (green), 0.6 (yellow), 0.7 (orange), 0.8 (red), 0.9 (magenta), and 0.95 (purple). Models at $Z=0.014$ ({\it left}), $Z=0.006$ ({\it centre}), and $Z=0.002$ ({\it right}).}
\label{HRD7}
\end{figure*}

The physical ingredients included in our models are the same as in \citetalias{Ekstrom2012a}, except for the following points:
\begin{itemize}
\item Contrary to \citetalias{Ekstrom2012a}, some of the models presented in the present paper evolve for a significant time near the critical velocity. To follow the evolution of the surface velocities as best as possible, as well as the total angular momentum content, the effect of anisotropic winds are accounted for as in \citet{Georgy2011a}.
\item Some of our models reach the critical velocity during their evolution. In that case, the mechanical equatorial mass loss is accounted for as described above.
\end{itemize}

\begin{figure*}
\begin{center}
\includegraphics[width=\textwidth]{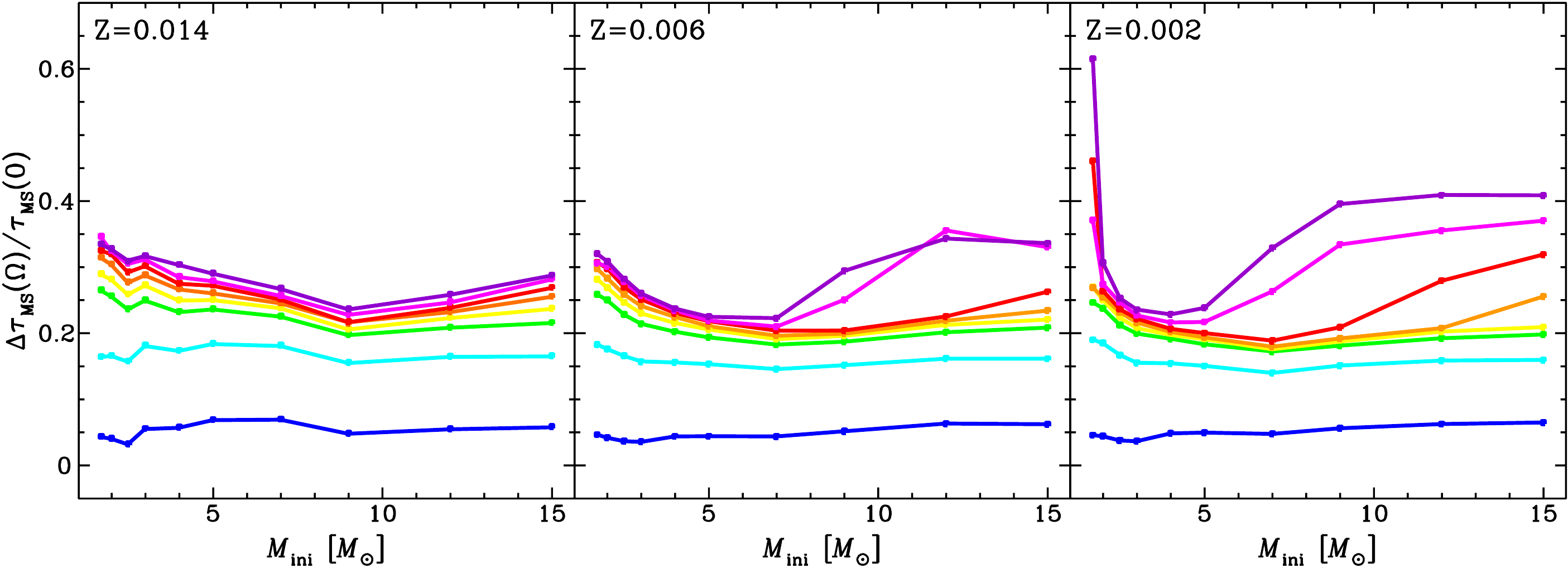}
\end{center}
\caption{MS lifetime enhancement as a function of the initial mass for all the models (same colour code as in Fig.~\ref{HRD7}). Models at $Z=0.014$ ({\it left}), $Z=0.006$ ({\it centre}), and $Z=0.002$ ({\it right}).}
\label{dtMS}
\end{figure*}

Our grid of models contains ten different masses ($1.7$, $2.0$, $2.5$, $3.0$, $4.0$, $5.0$, $7.0$, $9.0$, $12.0$ and $15.0\, M_{\sun}$), with nine different initial rotation rates ($\omega = 0$, $0.1$, $0.3$, $0.5$, $0.6$, $0.7$, $0.8$, $0.9$, $0.95$), and this for three different metallicities: $Z=0.014$ (solar metallicity), $Z=0.006$ (Large Magellanic Cloud metallicity) and $Z=0.002$ (Small Magellanic Cloud metallicity). This grid thus covers a mass range centred on the B-type stars\footnote{\footnotesize{As in \citetalias{Granada2012a}, we consider that a star is a B-type star if it has an effective temperature such as $10000\,\text{K} < T_\text{eff} < 30000\,\text{K}$}}. The models are evolved up to the helium flash ($M_\text{ini} \le 2\, M_{\sun}$), the early asymptotic giant branch ($2.5\, M_{\sun} \le M_\text{ini} \lesssim 9\, M_{\sun}$) or the end of central carbon burning ($M_\text{ini} \gtrsim 9\, M_{\sun}$).

For the non-solar metallicities, the initial abundances are set as follows
\begin{itemize}
\item The hydrogen and helium abundances are computed assuming that they vary linearly with respect to the total metal content $Z$ from the values given by the Big Bang nucleosynthesis to the present values. We thus have:
\begin{equation}
Y(Z) = Y_\text{P} + \frac{\Delta Y}{\Delta Z}Z,
\end{equation}
with $Y_\text{P} = 0.2484$ the primordial helium abundance \citep{Cyburt2003a}, and $\Delta Y/\Delta Z  = 1.257$. The hydrogen mass fraction is given by $X=1-Y-Z$.
\item The relative abundances of all other chemical species are the solar ones (the abundances are only scaled on a lower total metal content).
\end{itemize}

\subsection{Note on the diffusion coefficients and the calibrations}

Stellar evolution computations with one-dimensional codes such as those presented in this paper account for physical processes that are not unidimensional (convection, rotation, turbulent mixing, \textit{etc.}). One accounts for these effects through simplified theories, which have various free parameters that have to be calibrated on observations: overshoot parameter, mixing length, and shear mixing efficiency. Moreover, the implementation of the rotation includes two diffusion coefficients: the horizontal one $D_\mathrm{h}$ and the shear one $D_\mathrm{shear}$. In the literature, there are three different expressions for $D_\text{h}$ \citep{Zahn1992a,Maeder2003a,Mathis2004a} and two for $D_\mathrm{shear}$ \citep{Maeder1997a,Talon1997a}. Combining them allows for six different implementations of the rotation, and there is to date no reason to prefer any one of them on the basis of theoretical considerations.

The calibration of the overshoot parameter and of the mixing length is detailed in \citetalias{Ekstrom2012a}. In the framework of this paper, we add some complementary details on the choice and the calibration of the diffusion coefficients related to the rotation. In this work, as well as in \citetalias{Ekstrom2012a}, we chose $D_\mathrm{h}$ as in \citet{Zahn1992a} and $D_\mathrm{shear}$ as in \citet{Maeder1997a}. This choice relies mainly on it is allowing the following two behaviours simultaneously:
\begin{itemize}
\item The rotating models with an initial rotational velocity that we assume to be the more representative \citep[$v_\text{eq, ini}/v_\text{crit} = 0.4$, based on the work of][see also \citetalias{Ekstrom2012a} for more details]{Huang2006a} have a blue loop in the mass range $\sim 5 - \sim 9\, M_{\sun}$;
\item The mixing at the edge of the convective core during the MS is strong enough, leading to a bigger core than in the non-rotating case. This allows rotation to increase the size of the convective core in agreement with recent asteroseismic observations by \citet{Neiner2012a}, who support the view that rotation enlarges the convective core.
\end{itemize}

The efficiency of the mixing \citepalias[the $f_\text{energ}$ factor in eq.~(4) in][]{Ekstrom2012a} was calibrated to reproduce typical chemical enrichments at the surface of solar-metallicity MS B-type stars for our assumed typical initial rotation velocity \citepalias[the rotating models in][see their Fig.~11]{Ekstrom2012a}. To be consistent with the previous work, we kept the same value for all the models presented here.

\subsection{Electronic data}

All the electronic tables of these models are available at \url{http://obswww.unige.ch/Recherche/evol/-Database-}, as well as at the CDS. We have developed an interactive web application allowing for the downloading of a stellar model with given $Z$, $M$, and $\Omega/\Omega_\text{crit}$ interpolated between the computed models presented here. It also offers isochrones computation at the desired age or age range. This application is available at the url \url{http://obswww.unige.ch/Recherche/evoldb/index/}. The database contains the models presented in this paper, as well as the models from \citet{Ekstrom2012a} and \citet{Mowlavi2012a}. In the near future, it will be extended with additional metallicities.

\begin{figure*}
\begin{center}
\includegraphics[width=.45\textwidth]{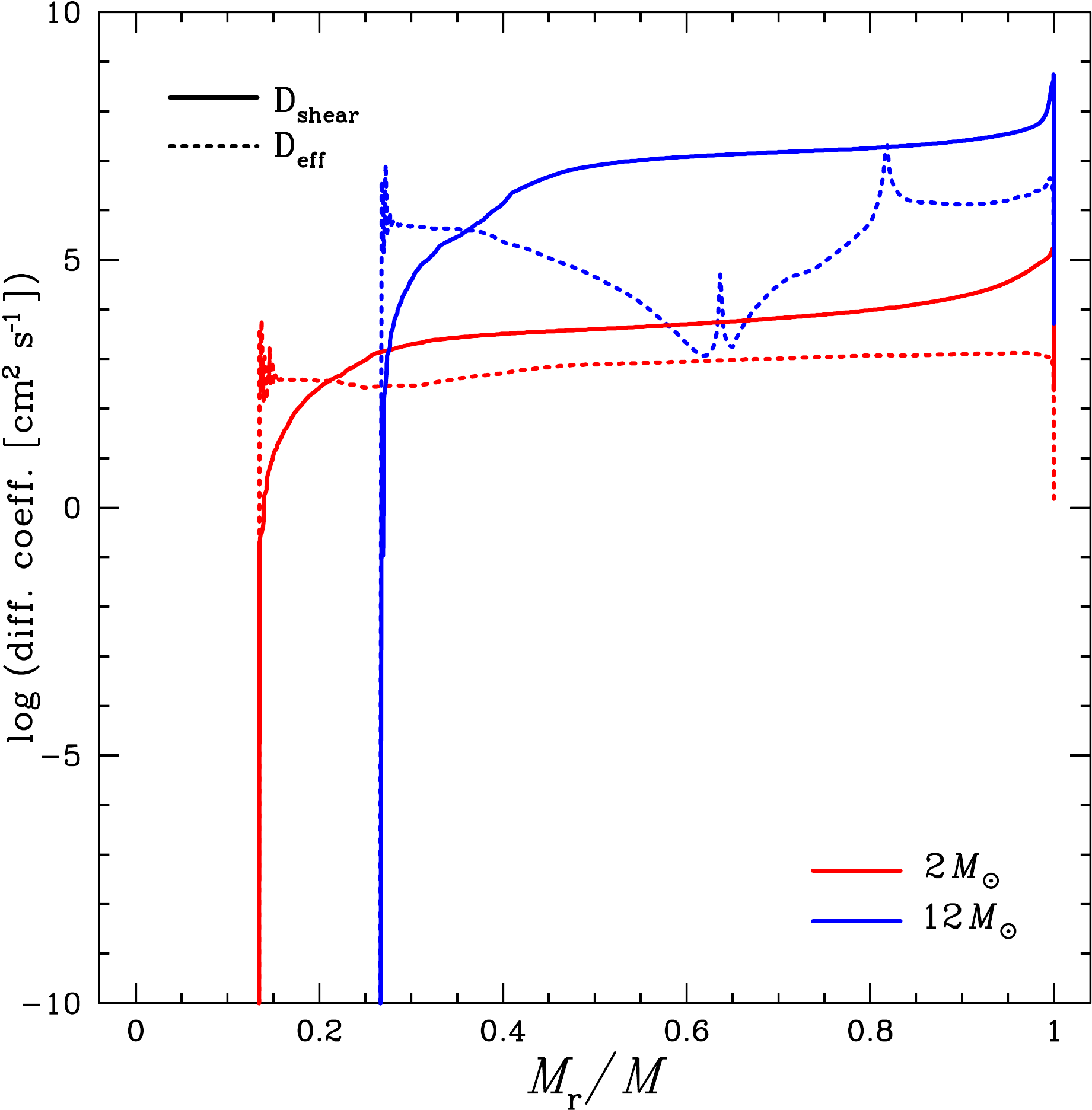}\hspace{.05\textwidth}\includegraphics[width=.45\textwidth]{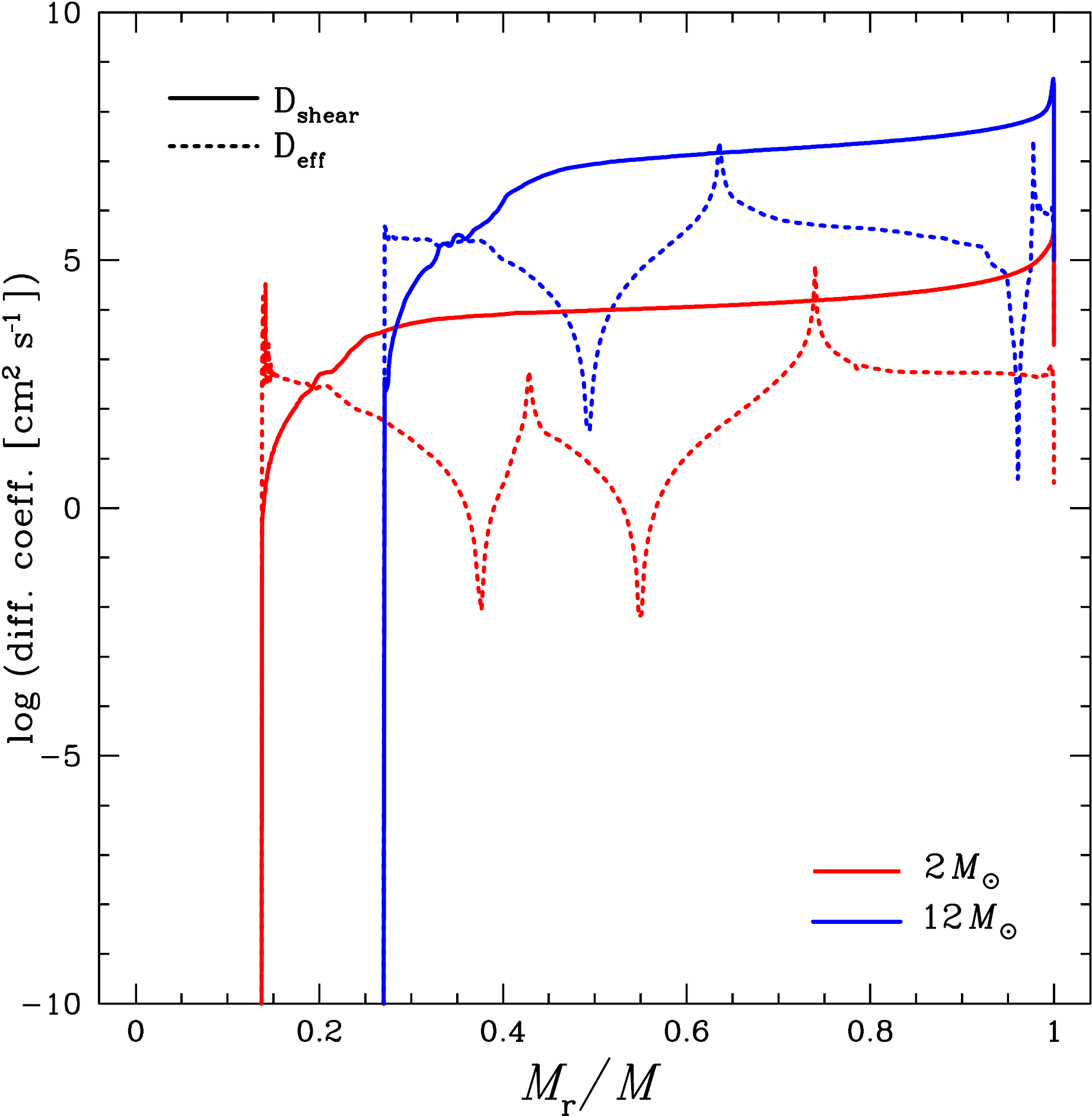}
\end{center}
\caption{Diffusion coefficients profiles for a $2$ (red curves) and a $12\, M_{\sun}$ model (blue curves) at $Z = Z_{\sun}$ \textit{(left panel)} and $Z=Z_\text{SMC}$  \textit{(right panel)}. The effective diffusion coefficient $D_\text{eff}$ (accounting for the effects of the meridional circulation on the chemical species) is plotted in dashed lines, and the shear mixing diffusion coefficient in solid lines.}
\label{MixEfficiency}
\end{figure*}

\begin{figure*}
\begin{center}
\includegraphics[width=.45\textwidth]{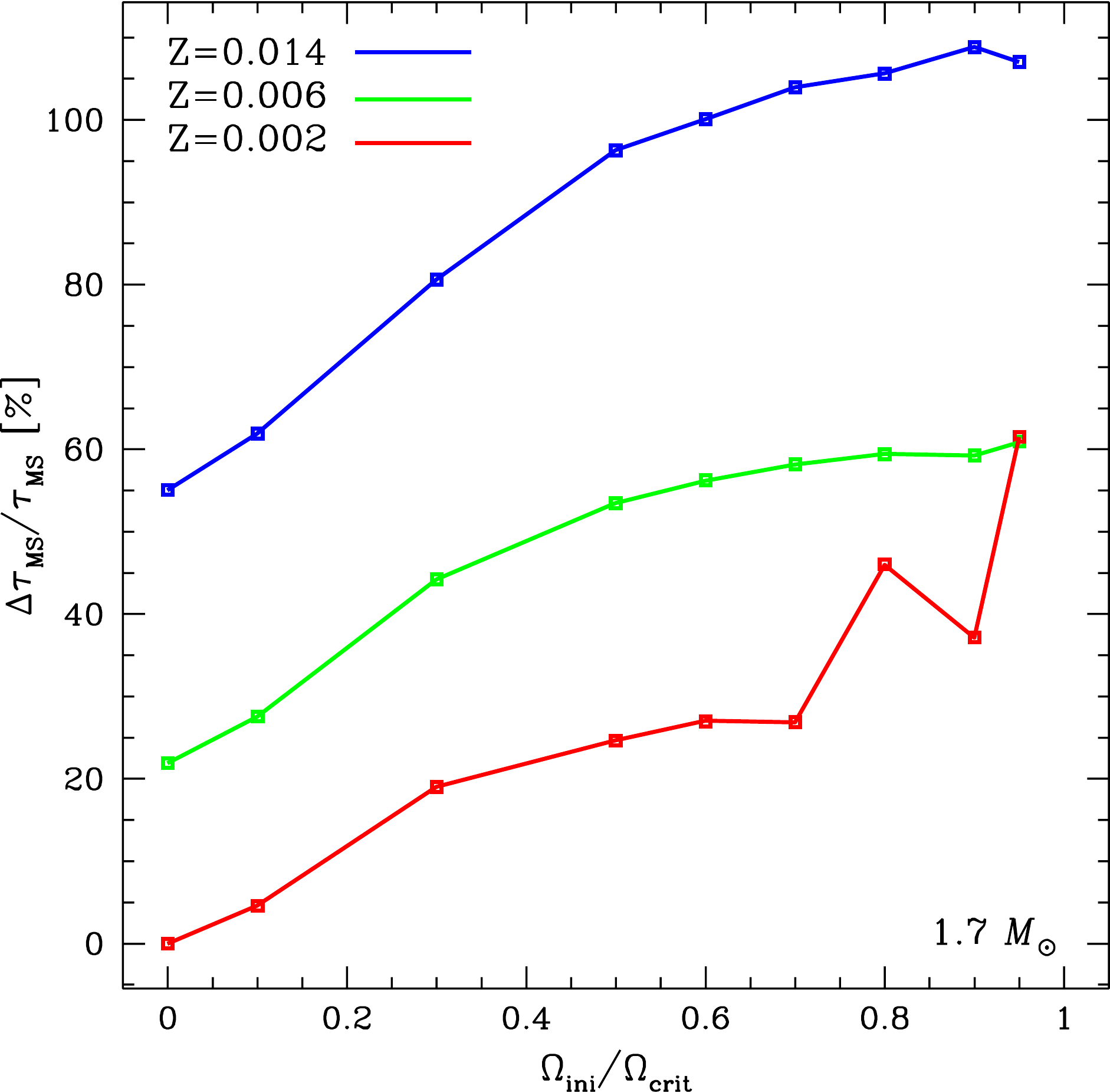}\hspace{.05\textwidth}\includegraphics[width=.45\textwidth]{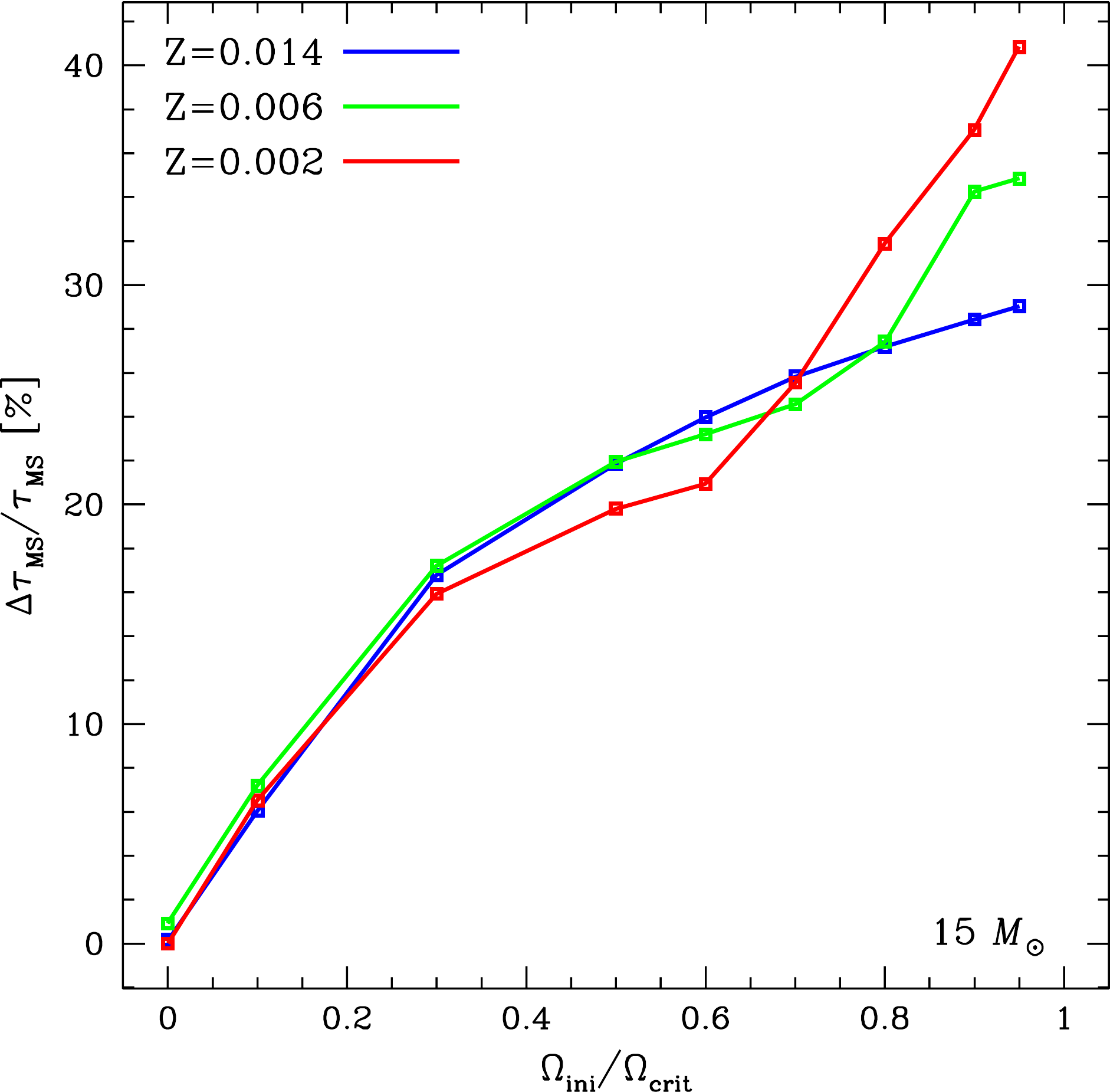}
\end{center}
\caption{MS duration enhancement as a function of the initial rotation rate. {\it Left:} 1.7 $M_{\sun}$ models. {\it Right:} 15 $M_{\sun}$ models.}
\label{tauOmega}
\end{figure*}

\section{Results \label{Results}}

The overall characteristics at the end of the main burning stages for all our models are presented in Tables~\ref{TabListModelsZ014} to \ref{TabListModelsZ002}. General Hertzsprung-Russell diagrams (HRD) with a colour scale indicating $\omega$ are presented in Figs.~\ref{HRD_Z014} to \ref{HRD_Z002}. The shaded area represents the approximative position of the Cepheid instability strip, according to \citet{Tammann2003a}.

\subsection{Mixing efficiency\label{SecMixing}}

In Fig.~\ref{MixEfficiency}, we show, for two masses and two metallicities, the profile of the diffusion coefficients responsible for the transport of the chemical species: the effective diffusion coefficient  $D_\text{eff}$ accounting for the composed effects of the meridional circulation and of the strong horizontal turbulence \citep[see][]{Chaboyer1992a}, and the shear-mixing diffusion coefficient $D_\text{shear}$, at roughly the middle of the MS phase ($X_\text{H, cen} = 0.3$). We see that except near the edge of the convective core, where $D_\text{eff}$ dominates, the shear mixing is responsible for the transport of the chemical species in most of the radiative envelope. We see a clear trend to comparing the two different masses for the same metallicity: both $D_\text{eff}$ and $D_\text{shear}$ are higher for a higher initial mass. In contrast, the coefficients do not show any noticeable difference with respect to the metallicity. However, due to the lower opacities at lower metallicity, low-$Z$ stars are more compact, leading to a shorter characteristic timescale for the diffusion ($\tau_\text{diff}\sim\frac{R^2}{D}$), and thus a more efficient enrichment of the surface (see the discussion in Sect.~\ref{SecChemicals}). This behaviour remains true for all the models at the three considered metallicities.

\subsection{HRD, lifetimes \label{SecHRD}}

As described by \citet{Meynet2000a} and \citet{Ekstrom2008b}, the position on the ZAMS is governed by the centrifugal force: rotating models behave like lower mass ones, with a shift in the tracks toward lower $L$ and $T_\text{eff}$. When the evolution proceeds, rotational mixing brings fresh hydrogen into the core, slowing down its decrease in mass. Also, newly produced helium is brought into the radiative zone, so the model evolves on a bluer and more luminous track. 

As shown in Fig.~\ref{HRD7}, the luminosity attained at the end of the MS, and thus during the crossing of the Hertzsprung-Russell gap, increases with the rotation rate up to $\omega_\text{ini}\leq0.5$. Beyond, the luminosity decreases again and the crossing occurs at a luminosity that is lower than the one of the $\omega_\text{ini}=0.3$ model, except at $Z=0.002$, where the luminosity at the end of the MS increases again above $\omega_\text{ini} \sim 0.8$. This behaviour is directly related to the size of the convective core, which governs the luminosity at the end of the MS. During the MS, the evolution of the size of the convective core relies on two counteracting physical processes linked to rotation. First, rotation generates an additional support against gravity due to the centrifugal force. These effects tend to decrease the size of the core and its luminosity. Second, the rotational mixing at the edge of the core progressively brings fresh material into the core, increasing its mass, hence its luminosity. Both effects are responsible for the non-monotonic behaviour of the luminosity at the HRD-crossing as a function of the initial rotation rate.

For the models going through a blue loop during core He-burning (passing through a classical Cepheid phase), we note that rotation increases the luminosity at which the loop occurs. The effect of rotation on the blue loops becomes spectacular at low metallicity, with the disappearance of the typical loop pattern for the most rapid rotators. We discuss this point in more detail in Sect.~\ref{SSLoops}.

Rotational mixing increases the MS lifetimes, as shown in Fig.~\ref{dtMS}. At solar metallicity, for the mass range considered, the increase does not much vary as a function of the initial mass. It amounts to about $15-25\%$ for $\omega_\text{ini}$ between 0.30 and 0.50. For the most rapid rotators considered here, it may reach values as high as $30-35\%$.

When the metallicity decreases, the variation with the initial mass becomes more marked at the low and high mass ranges considered. As a numerical example, at $Z=0.002$, the maximum increase reaches values as high as $62\%$ for the $1.7\,M_{\sun}$ and as high as 41\% for the $15\,M_{\sun}$.

\begin{figure*}
\begin{center}
\includegraphics[width=.45\textwidth]{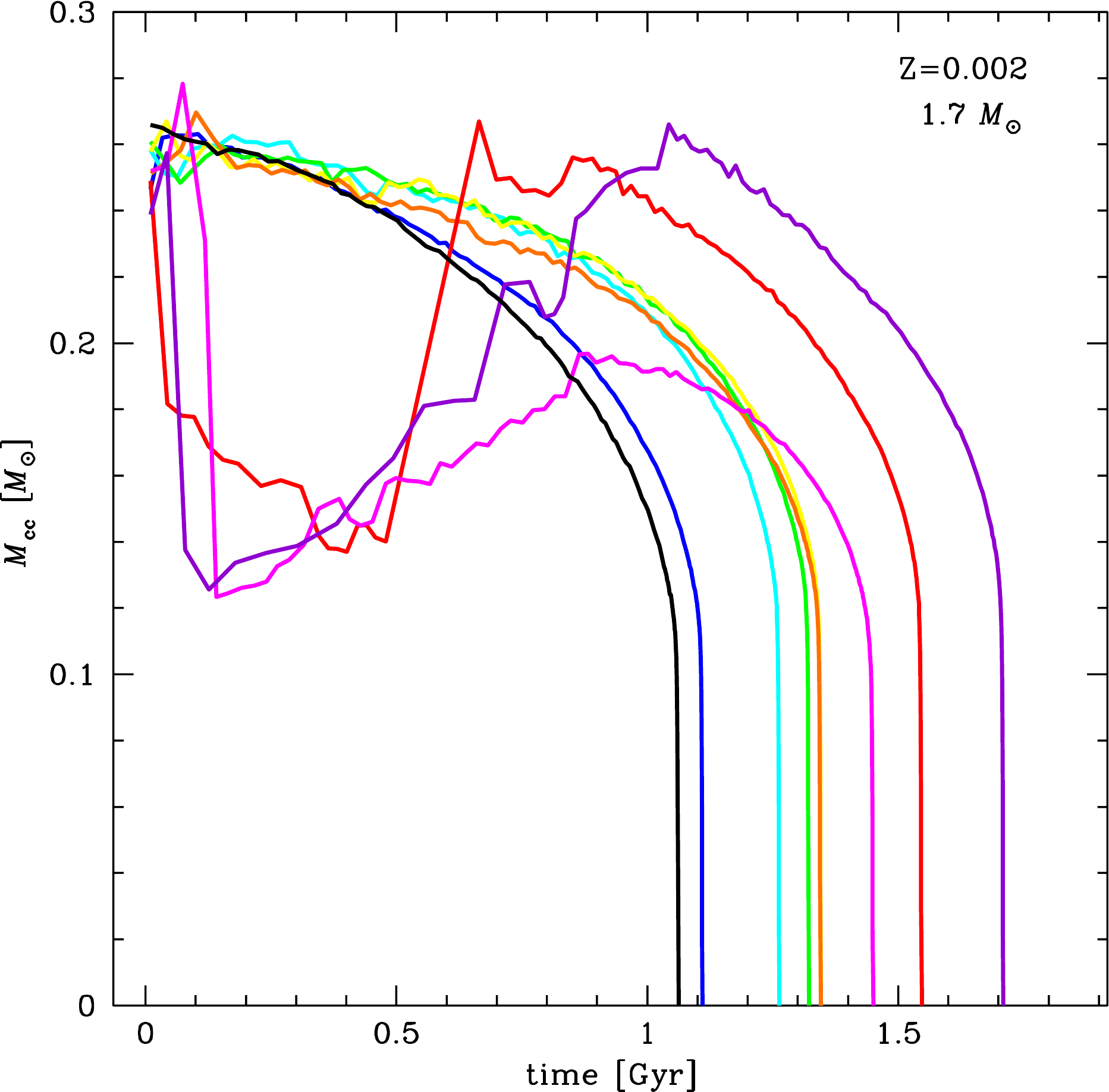}\hspace{.05\textwidth}\includegraphics[width=.45\textwidth]{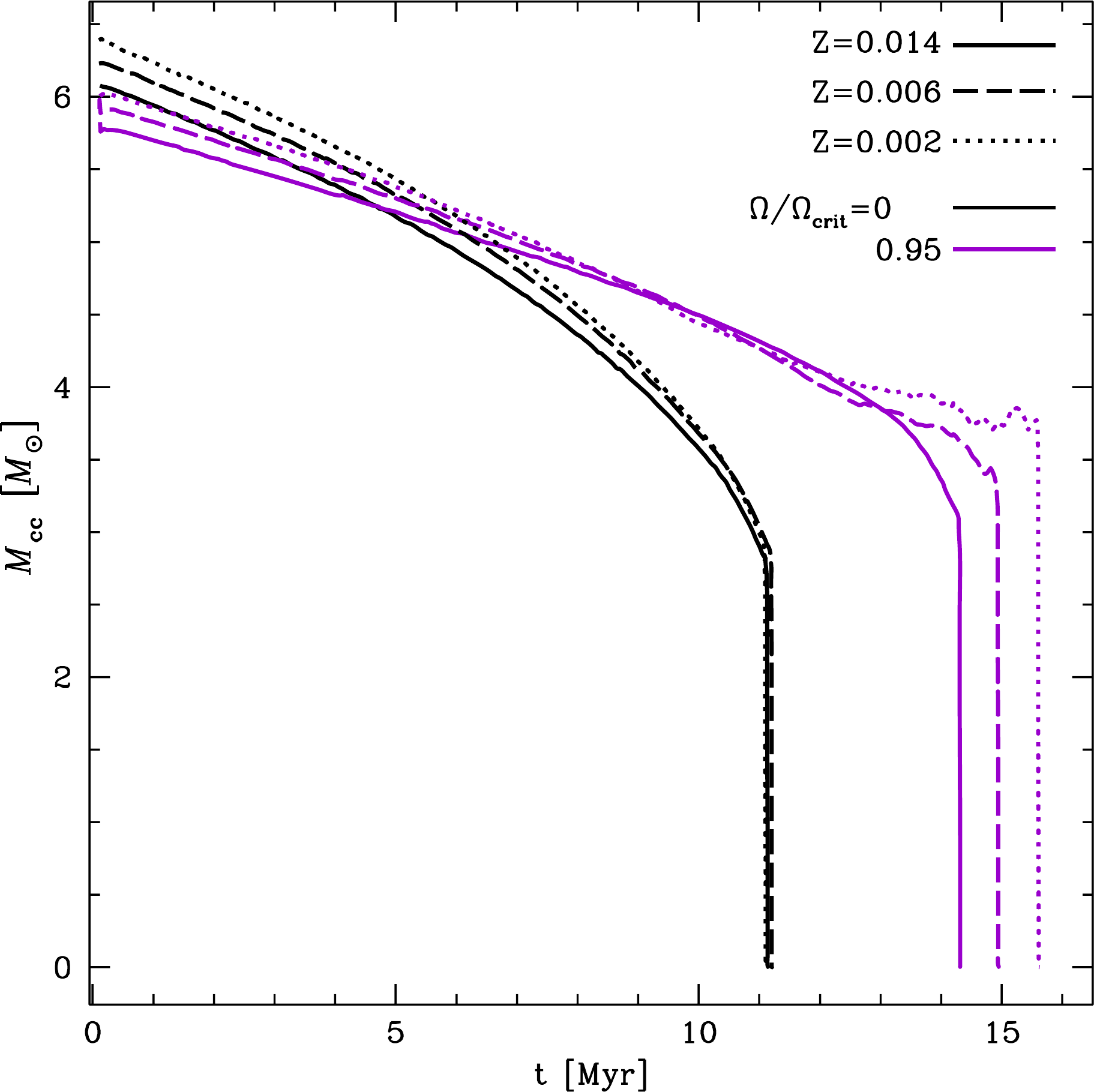}
\end{center}
\caption{Core mass evolution. {\it Left:} 1.7 $M_{\sun}$ models at $Z=0.002$ (same colour code as in Fig.~\ref{HRD7}). {\it Right:} 15 $M_{\sun}$ models at the three $Z$, and $\Omega/\Omega_\text{crit}=0$ and 0.95.}
\label{Mcc1p7-15}
\end{figure*}

The high sensitivity at $1.7\,M_{\sun}$ comes from the central temperature at the beginning of the core H-burning phase being around $20\,\text{MK}$, \textit{i.e.} in the range of temperature at which CNO-burning begins to dominate the pp-chain for the energy production. When rotation is fast enough, the hydrostatic effects of rotation decrease the central temperature, and therefore the pp-chain becomes the dominant energy production channel. Typically, at $200\, \text{Myr}$, the three more rapidly rotating models have $T_\text{c}<21.4\, \text{MK}$, while all the others have $T_\text{c}>21.7\, \text{MK}$. As a result, the convective core is smaller during the first part of the MS phase (see Fig.~\ref{Mcc1p7-15}, \textit{left})\footnote{The decrease shown in Fig.~\ref{Mcc1p7-15} (\textit{left}) is of the same order of magnitude as the one obtained by \citet{Mowlavi2012a} when the initial mass decreases from $1.7$ down to $1.6\,M_{\sun}$, precisely for the same effect as on the energy production channel.}. It is amazing to note that such a slight difference in $T_\text{c}$ is enough to pass from pp-chain to CNO-cycle dominated energy production. When evolution proceeds, and the central temperature increases, the convective core grows and reaches values even higher than the values obtained at the beginning of the core H-burning phase in the non-rotating model.

At low metallicities and masses above $5-7\,M_{\sun}$, we obtain large increases of the MS lifetimes for $\omega_\text{ini}>0.70$. Actually, some small convective zones develop in the region above the core, where there is a variable chemical composition. (This phenomenon is related to the phenomenon of  ``semi-convection''.) These convective zones mix the region above the core (and, in some cases, even merge with the core), bringing fresh hydrogen close to it, favouring a refuelling of the core, and in turn increasing the lifetime on the MS. The details in the development of these intermediate convective zones are very sensitive to the precise prescription used for rotation, however the general features seem robust and occur independently of the physical ingredients.

In Fig.~\ref{tauOmega}, we compare the degree of dispersion of the MS lifetimes due to change of metallicity and of rotation for two initial masses. We see that for the $15\,M_{\sun}$ (Fig.~\ref{tauOmega}, \textit{right}), up to $\omega_\text{ini}=0.60$, the dispersion due to rotation is much greater than the one due to metallicity. This is due to the change in metallicity for the hot massive stars that have only moderate effects on the opacities\footnote{The opacity in the temperature range of the massive stars is dominated by electron scattering, which is only marginally affected by a change in the chemical composition.}, hence on the luminosity of the stars. In that case, the mixing is the dominant effect changing the MS duration. At high velocities, however, the change brought by varying the metallicity are more important, but this is again a consequence of rotational mixing whose consequences are not the same depending on the initial metallicity (see above).

In the low mass-range ($1.7\,M_{\sun}$, Fig.~\ref{tauOmega}, \textit{left}), we somehow have an inverse situation, in the sense that, in general, the effects of changing the metallicity have a greater impact on the MS lifetime than changing the initial velocity. This reflects the fact that, on the one hand, rotational mixing becomes less efficient in low-mass stars (see Fig.~\ref{NHendH}), and on the other hand, the impact of a change in $Z$ on the opacity is much greater than in the higher mass range.

\subsection{Core mass vs rotation and metallicity}

As mentioned above, the additional support against gravity produced by the rotation allows the central temperatures to be lower for increasing initial velocities. This implies that the size of the convective core on the ZAMS decreases as a function of the initial rotation velocity. However, as evolution proceeds, the more efficient mixing in the more rapidly rotating stars increases the size of the convective core, which becomes bigger than the core of more slowly rotating stars. At the end of the MS, the faster the stars initial rotation, the larger the convective core. This behaviour is valid at the three considered metallicities. We recall, however, the peculiar behaviour of the convective core of the $1.7\, M_{\sun}$ at different rotational velocities, owing to the conflicts between the pp-chains and CNO cycle (see above and Fig.~\ref{Mcc1p7-15}, left panel).

At low metallicity, for the upper mass range of our sample (above $\sim7\, M_{\sun}$), and for the fastest rotational velocities ($\omega \gtrsim 0.8$), the merger of the core with the convective zone that develops above it modifies the behaviour at the end of the MS, refuelling it in fresh hydrogen and prolonging the MS duration (see Fig~\ref{Mcc1p7-15}, right panel). The core mass will have an effect on the final yields, as is discussed in Sect.~\ref{SSEject}.

\subsection{Evolution of the surface velocity}

\begin{figure*}
\begin{center}
\includegraphics[width=.45\textwidth]{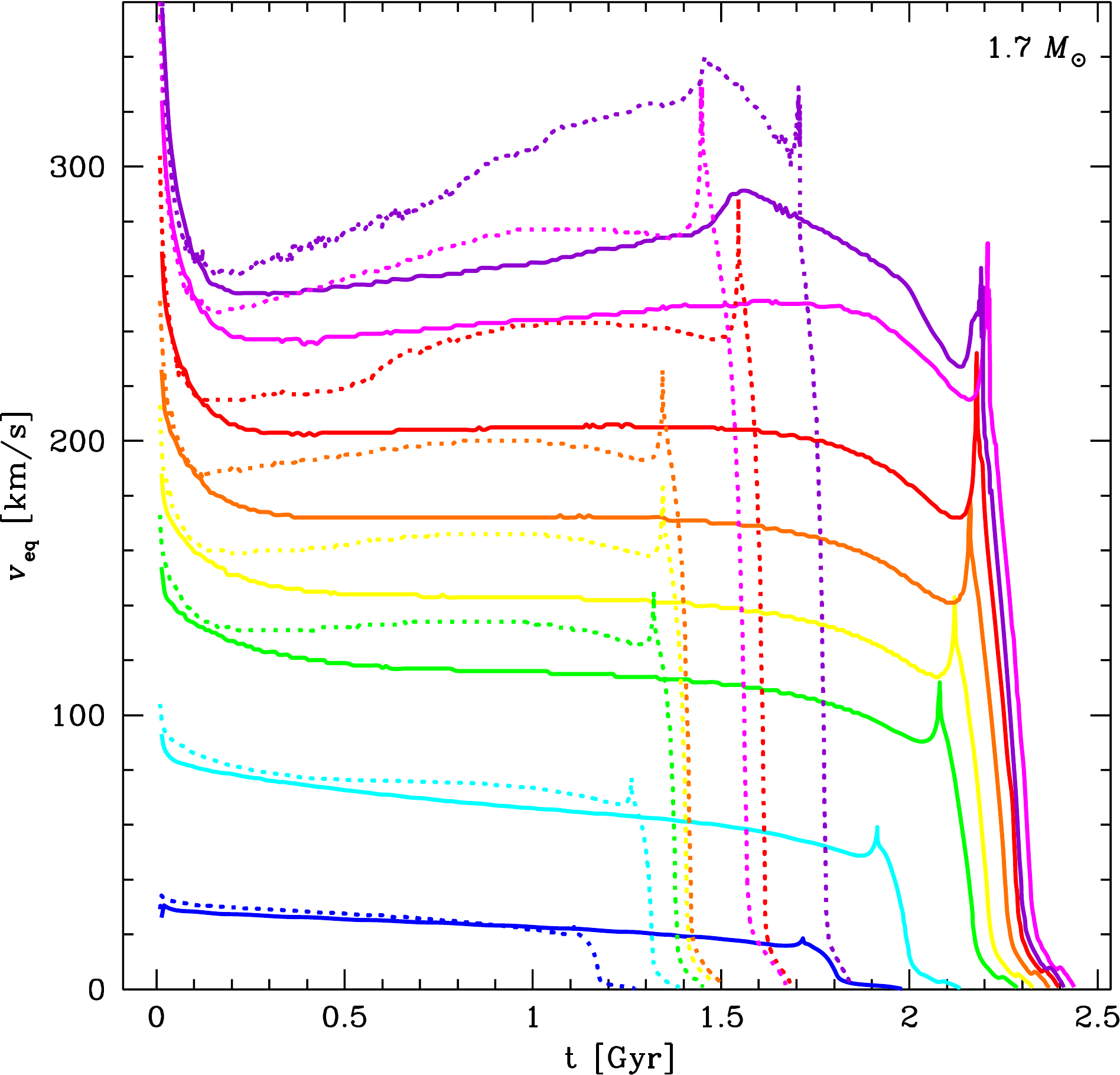}\hspace{.05\textwidth}\includegraphics[width=.45\textwidth]{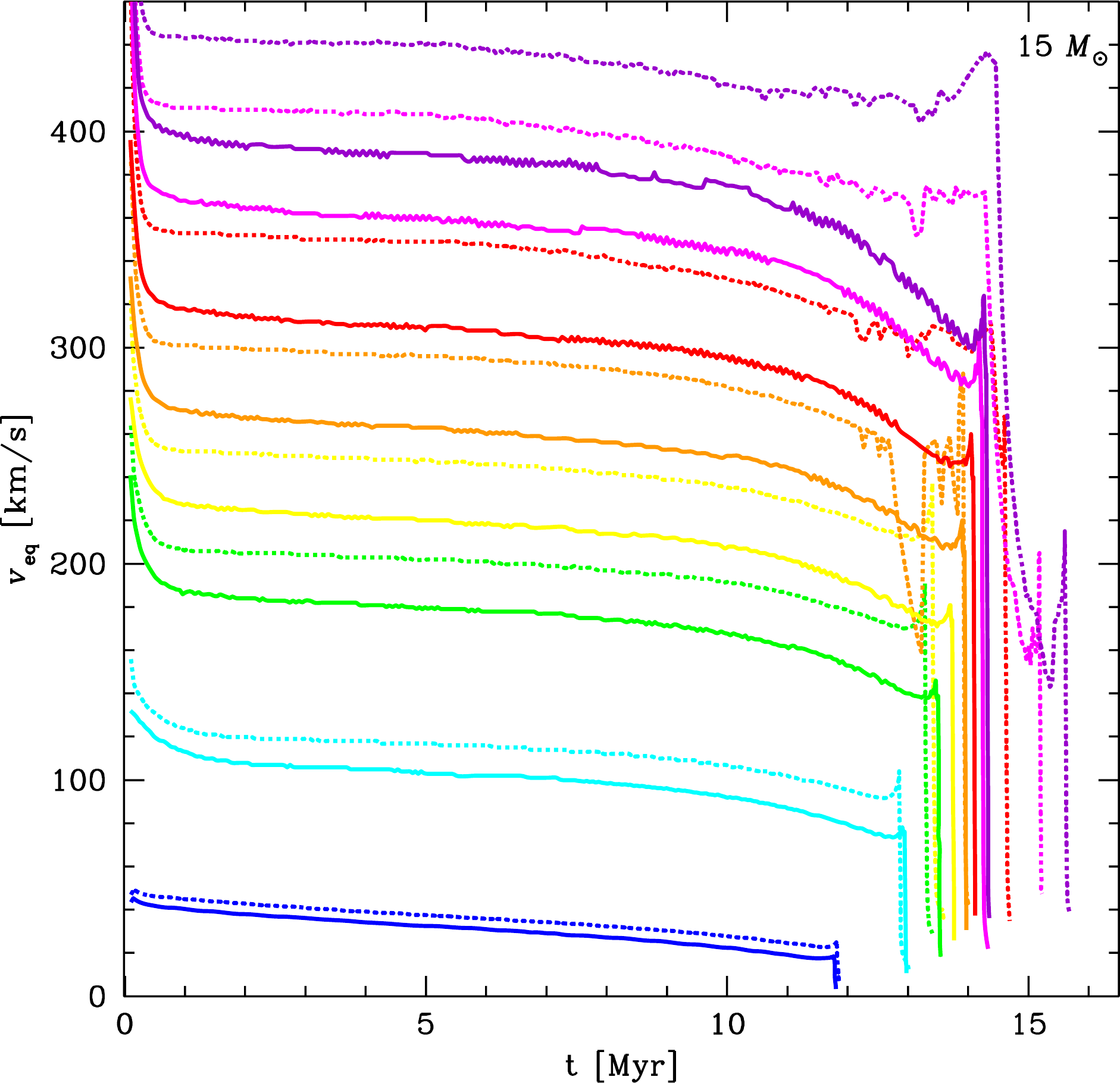}
\end{center}
\caption{Equatorial velocity evolution (same colour code as in Fig.~\ref{HRD7}) at $Z=0.014$ (solid lines) and $Z=0.002$ (dotted lines). \textit{Left:} $1.7\,M_{\sun}$ models. \textit{Right:} $15\,M_{\sun}$ models.}
\label{Vsurf}
\end{figure*}

Figure~\ref{Vsurf} shows the evolution of the surface velocities during the MS phase for the $1.7$ and $15\,M_{\sun}$ stellar models. Close to the ZAMS, we see an abrupt decrease in $v_\text{eq}$. It corresponds to the time taken by the model to reach a quasi-equilibrium state (the initial flat $\Omega$-profile is not at the equilibrium once the meridional circulation acts). We also see that the equatorial velocities are initially higher at lower metallicity for the models starting at the same $\omega$. This is due to the larger compactness of the star at low $Z$: since the initial CNO content is lower, the star must compensate for this lack of catalyst elements during the core hydrogen-burning phase by adopting a more compact structure that allows a higher central temperature. With $v_\text{eq, ini} = \Omega_\text{ini}r$, and looking at eq.~\ref{EqOmegaCrit}, we find that for a given $\omega$ and a given mass, the initial equatorial velocity is $v_\text{eq, ini} \sim r^{-\frac{5}{2}}$, and thus, is higher for a more compact star.

The surface angular velocity varies as a function of the time because of three main physical processes:
\begin{itemize}
\item the local conservation of the angular momentum, which modifies the angular velocity when the star contracts or expands;
\item internal transport mechanisms \citep[see][]{Zahn1992a,Maeder1998a}, which redistribute the angular momentum throughout the star;
\item stellar winds, which remove angular momentum from the stellar surface.
\end{itemize}
During the MS, our models develop an external meridional circulation cell \citep[Gratton-\"Opik cell, see \textit{e.g.}][]{Maeder2009a}, which carries angular momentum from the inner part of the star to the surface, tending to accelerate it. The efficiency of the meridional circulation is greater for higher masses, for higher initial rotational rates and for higher metallicity. In contrast, the stellar mass loss due to radiative winds \citep{Castor1975a} tends to brake the stellar surface. As for the meridional circulation, the strength of the stellar winds is greater for higher mass stars, for higher rotational rates \citep{Maeder2000a}, and for higher metallicities. There are thus two counteracting effects governing the evolution of the stellar surface velocity. For a given initial mass, rotation rate, and metallicity, their relative efficiency will govern the increase or decrease in the surface velocity during stellar life. We see from Fig.~\ref{Vsurf} that in the mass range studied here, the equatorial velocity remains roughly constant during the MS, however, according to Eq.~\ref{EqVcrit}, $v_\text{crit}$ itself decreases during the MS, since $R_\text{eq}$ is increasing. As a result, the ratio $v_\text{eq}/v_\text{crit}$ increases \citep[see][]{Ekstrom2008b}.

\begin{figure}
\begin{center}
\includegraphics[width=.45\textwidth]{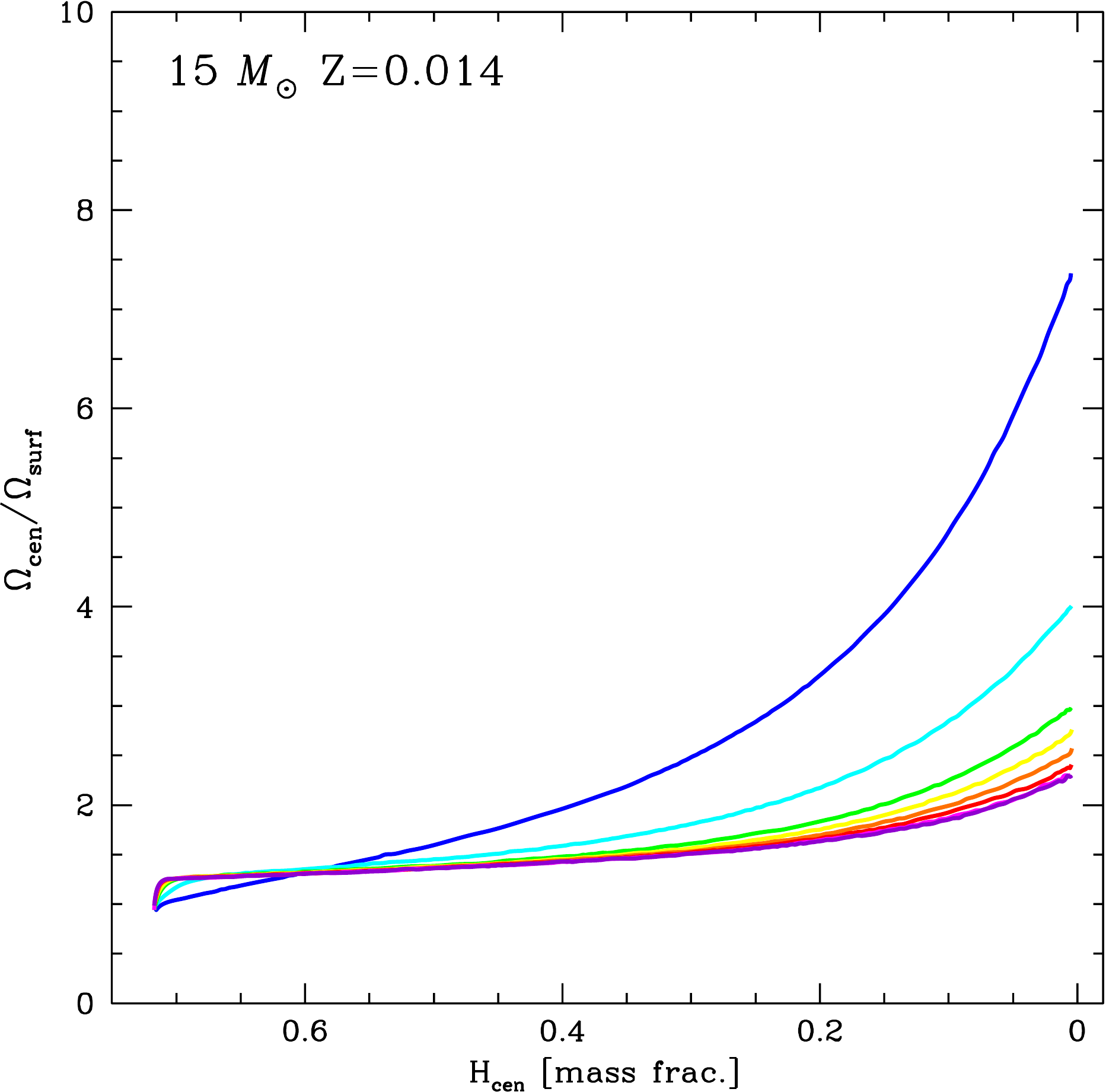}
\end{center}
\caption{$\Omega_\text{centre}/\Omega_\text{surface}$ ratio for the $15\,M_{\sun}$ models et $Z=0.014$ (same colour code as in Fig.~\ref{HRD7}).}
\label{OcOs015}
\end{figure}

From Fig.~\ref{OcOs015}, one sees that the contrast between the rotation rates of the core and of the envelope is stronger in slowly rotating stars than in faster rotating ones. This illustrates that the combined effects of meridional circulation and shears, which tend to flatten the $\Omega$-profiles, are stronger in faster rotating stars. Similar qualitative behaviours are obtained for the whole mass and metallicity range considered here. The model with $\omega_\text{ini}=0.50$ \citep[which is close to the peak of the rotation rate distribution according to][]{Huang2010a} shows a ratio $\Omega_\text{cen}/\Omega_\text{surf}\simeq3$. At the moment, the ratio of $\Omega_\text{cen}/\Omega_\text{surf}$ has been estimated through asteroseismology in three slowly rotating B-type stars \citep{Aerts2008a}. The results range between 1 and 6. The present results overlap the observed range well, although the small amount of data does not allow a systematic and detailed comparison.

\subsection{Surface abundances}\label{SecChemicals}
\begin{figure*}
\begin{center}
\includegraphics[width=\textwidth]{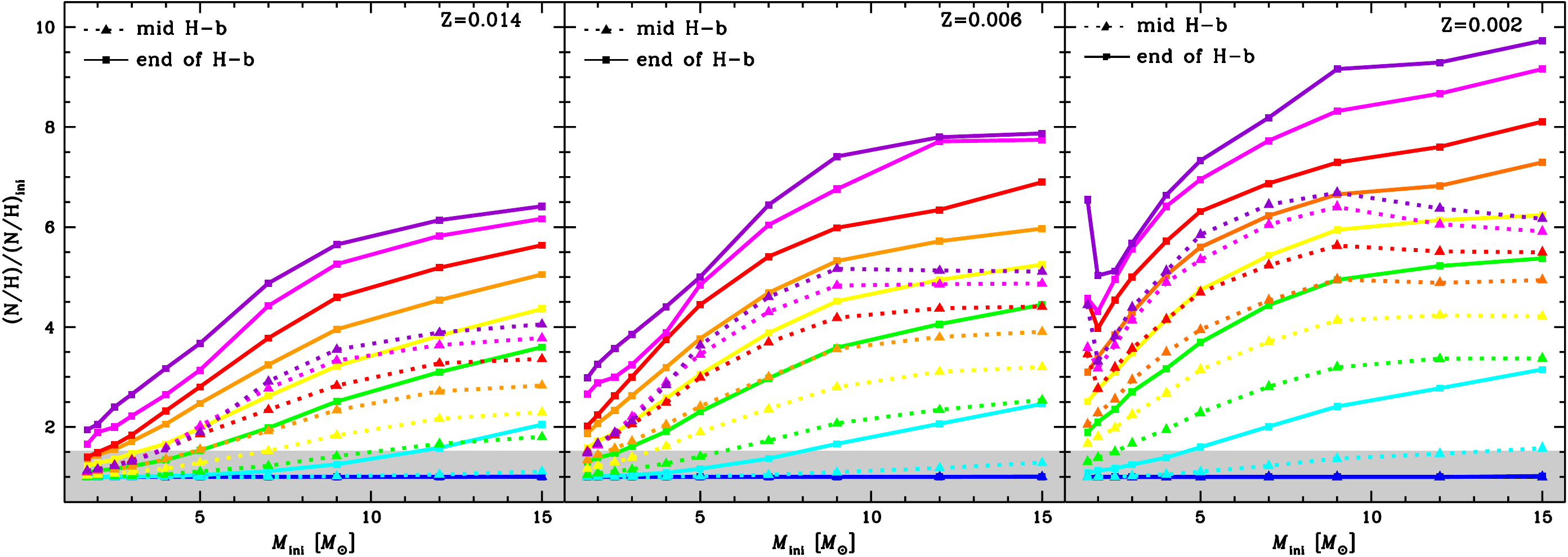}
\end{center}
\caption{N/H ratio at the end of the MS as a function of the initial mass for all the models. Models at $Z=0.014$ ({\it left}), $Z=0.006$ ({\it centre}), and $Z=0.002$ ({\it right}).}
\label{NHendH}
\end{figure*}

The variations in the surface abundances at the end of the MS phase are shown in Fig.~\ref{NHendH}. We see, as already obtained in previous works \citep{Maeder2001a}, that for a given value of $\omega_\text{ini}$\footnote{\footnotesize{We focus our discussion on comparing models with the same $\omega_\text{ini}$. The trends can differ slightly if we considered for example the same $v_\text{eq, ini}$. The detailed trends can be obtained by the mean of the tables provided in this paper, or directly from the available electronic data.}}, the enrichment is stronger at lower metallicities and for higher masses. This is true not only for the massive stars, but also for intermediate-mass stars.

In Fig.~\ref{NHendH}, the shadowed area corresponds to a variation of $0.2\,\text{dex}$ with respect to the initial N/H ratio (corresponding roughly to the typical error bars in the measurements of abundances. For example, typical error bars in the VLT-FLAMES survey of massive stars for individual stars are estimated by the authors to be between 0.1 and 0.3 dex \citep{Hunter2007a}. According to \citet{Nieva2010a}, the typical error bars are systematically underestimated, and should rather be of the order of 0.3 dex. The models evolving out of this area will therefore have a strong enough enrichment to be observable. The dashed lines correspond to the mid-MS (when the central hydrogen mass fraction is half of its initial value), and the continuous lines to the end of the MS. At solar metallicity, the mixing is efficient enough in massive stars ($M \ge 9\, M_{\sun}$) to be observationally detectable, even for moderately rotating ($\omega \gtrsim 0.5$) stars, and already at the middle of the MS. For the lower masses, this is no longer true, and only stars with an initial mass above $4\, M_{\sun}$ are mixed enough to produce an observable enrichment at the middle of the MS, even for the fastest rotators. At the end of the MS, the enrichment is observable only for the most rapidly rotating stars for our $1.7\, M_{\sun}$ models ($\omega \gtrsim 0.8$).

At $Z=0.002$, we see that the more efficient mixing makes the surface enrichment much more easily detectable, even for our lowest mass models. Indeed, all the stars with initial velocity $\omega_\text{ini} > 0.60$ (yellow tracks) show sufficient enrichment at mid-MS, and $\omega_\text{ini}  > 0.50$ for the end of the MS.

\section{Advanced phases\label{Sadvanced}}
\subsection{Blue loops\label{SSLoops}}

The physical mechanisms leading to the expansion or contraction of the envelope during the stellar life are still poorly understood, and are being debated \citep{Renzini1992a,Stancliffe2009b}. According to \citet{Lauterborn1971a}, the occurrence of a blue loop during core He-burning depends on the relation between the gravitational potential of the core and a critical potential that is mass dependent, as well as on the internal hydrogen profile. Any process able to modify either the total mass, the mass or radius of the core, or the hydrogen abundance profile will affect the occurrence of a blue loop.

\begin{figure*}
\begin{center}
\includegraphics[width=\textwidth]{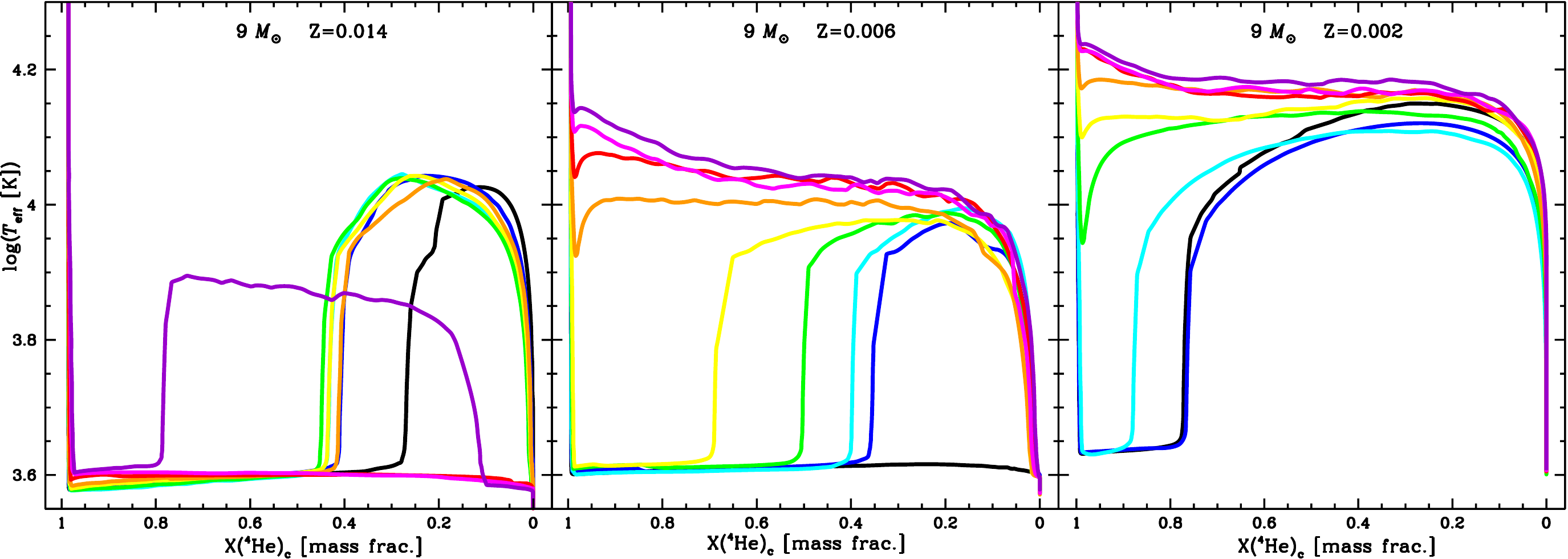}
\end{center}
\caption{$T_\text{eff}$ as a function of the helium mass fraction in the core. Models at $Z=0.014$ ({\it left}), $Z=0.006$ ({\it centre}), and $Z=0.002$ ({\it right}).}
\label{Loops9}
\end{figure*}

\begin{figure*}
\begin{center}
\includegraphics[width=.9\textwidth]{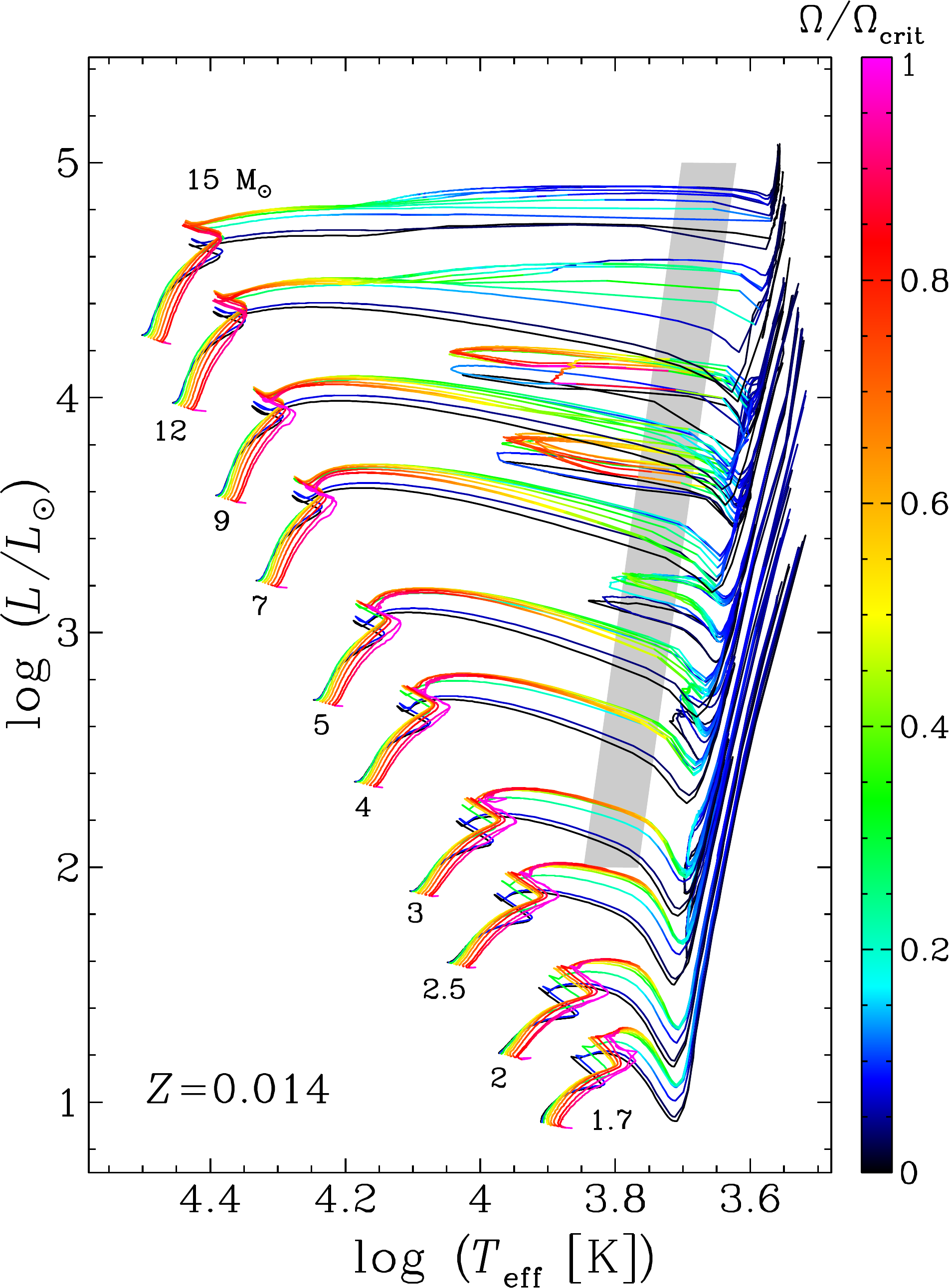}
\end{center}
\caption{HRD for $Z=0.014$. The colour code represents  the ratio $\Omega/\Omega_\text{crit}$ (scale on the right). For each mass, the order of the curves on the ZAMS(bottom-left point of the curves) goes from the lowest to the highest initial rotation velocity from left to right.}
\label{HRD_Z014}
\end{figure*}

\begin{figure*}
\begin{center}
\includegraphics[width=.9\textwidth]{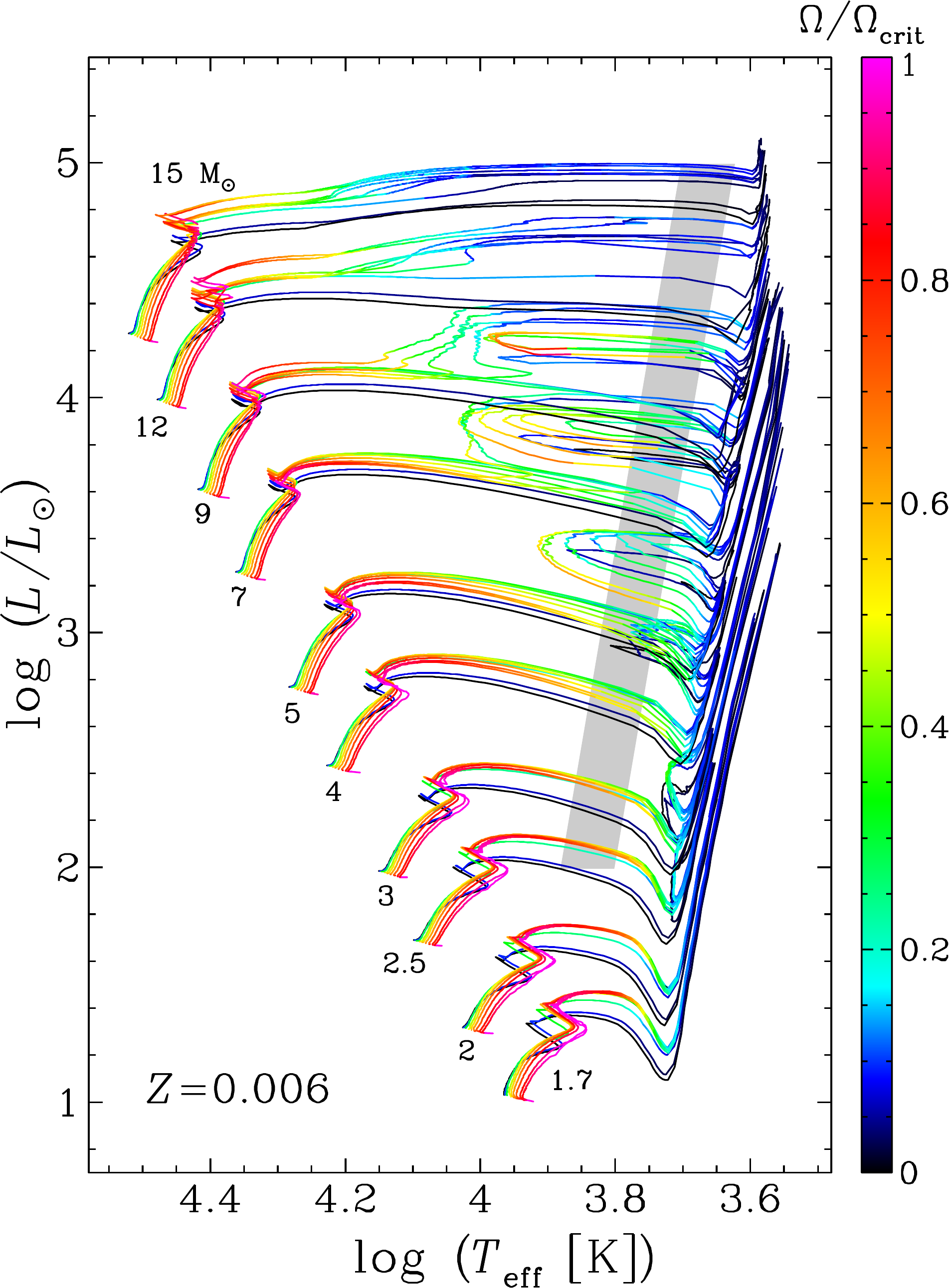}
\end{center}
\caption{Same as Fig.~\ref{HRD_Z014}, but for $Z=0.006$.}
\label{HRD_Z006}
\end{figure*}

\begin{figure*}
\begin{center}
\includegraphics[width=.9\textwidth]{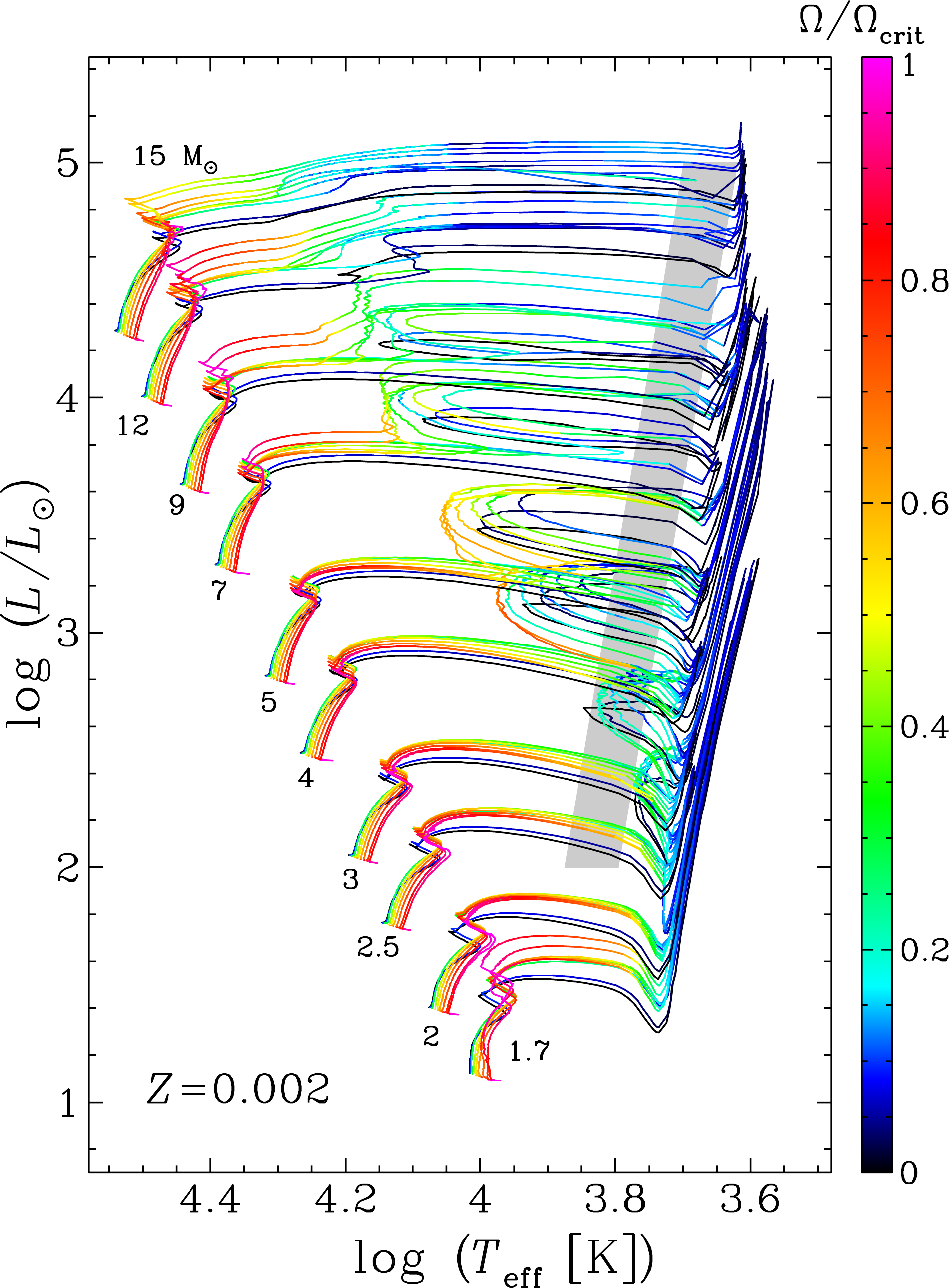}
\end{center}
\caption{Same as Fig.~\ref{HRD_Z014}, but for $Z=0.002$.}
\label{HRD_Z002}
\end{figure*}

Rotation affects all these quantities in various ways, depending on the rotation rate, as can be seen in Fig.~\ref{Loops9}. Metallicity also plays a role, since rotational mixing occurs differently at different metallicities. A general feature is that rotation increases the time spent on the loop. An exception to that are the models at $Z_{\sun}$ rotating with $\omega_\text{ini} = 0.8$ and $0.9$, where the loop is suppressed.

At non-solar metallicity, the difference in luminosity between the leftwards excursion at the start of the loop and the redwards return movement is widened by rotation. The widening occurs toward both lower starting luminosity and higher final luminosity. In the most extreme cases, the lowering of the starting luminosity might reach the luminosity of the crossing of the Hertzsprung gap. In that case, the model ``jumps'' directly onto the top of the loop, avoiding the first RSG phase. For those models, most of the core He-burning occurs thus in the blue part of the HRD ($\log (T_\text{eff}\ [\text{K}])\gtrsim 4.0$). Such behaviour will reduce the number of RSG at low metallicity.

We will dedicate a future paper to more detailed study of the implication of this effect on the Cepheid population predicted by our models.

\subsection{Mass ejections of various elements\label{SSEject}}

In Table~\ref{TabYields}, we present the ejected masses for H, He, C, N, O, and remaining metals, as well as the CO-core mass and the remnant mass \citep[computed as in][and \citealt{Georgy2009a}]{Hirschi2005a}. We see that rotation increases the mass of the CO core significantly, and this in a larger way at lower metallicity, which matches the result of a higher mixing in lower $Z$. For the SMC, the increase amounts to 43\%, while for the LMC and Galaxy it is 27\% and 25\%, respectively, for the models with $\omega_\text{ini} = 0.95$. This table is qualitatively representative of the trends obtained for the higher part of our mass range. We do not discuss here the ejected masses for the models of our sample that become AGB stars, since we have not followed these models through this phase.

The general trends concerning the ejected elements in rotating models are the following.
\begin{itemize}
\item Hydrogen is mostly depleted ($\sim$60\% in the SMC and LMC models, and 73\% in the Galactic models); this reflects the fact that more mass is processed by nuclear burning in rotating stars.
\item Carbon and oxygen tend to be produced more. Comparing the models with $\omega_\text{ini} = 0$ and $\omega_\text{ini} = 0.95$, the increase in the ejected mass of carbon amounts to a factor of 2.15 for the SMC models, 1.52 for the LMC, and 1.71 for the Galactic models. The increase in the ejected mass of oxygen amounts to a factor of 3.36 for the SMC, 2.52 for the LMC, and 2.40 for the Galaxy. In the Galactic case, the higher increase in C and O occurs for models with $\omega_\text{ini}=0.50$ and $0.60$, respectively, with a maximal increase by a factor of  2.06 and 3.11, respectively.
\end{itemize}
Helium, nitrogen and the remaining metals do not show any definite trend. One also notes that even at high rotation, we do not observe any production of primary nitrogen, in these models.

\begin{table*}
\caption{CO-core mass, remnant mass, and ejected masses in hydrogen, helium, carbon, nitrogen, oxygen, and remaining heavy elements for the 15 $M_{\sun}$ models at the three metallicities and all the rotation rates. All masses are expressed in $M_{\sun}$ units.}
\label{TabYields}
\begin{center}
\begin{tabular}{lccccccccc}
\hline
\hline
    \rule[-1.5mm]{0mm}{2mm}    & $\omega_\text{ini}$ & $M_\text{CO}$ & $M_\text{rem}$\tablefootmark{1} & $M^{\text{ej}}_{\text{H}}$ & $M^{\text{ej}}_{\text{He}}$ & $M^{\text{ej}}_{\text{C}}$ & $M^{\text{ej}}_{\text{N}}$ & $M^{\text{ej}}_{\text{O}}$ & $M^{\text{ej}}_{\text{other heavies}}$ \\
\hline
\rule{0mm}{3mm}$Z=0.002$ & 0.00 & 2.6961 & 1.4265 & 6.8329 & 5.0423 & 0.3638 & 0.0056 & 0.6237 & 0.2129 \\
        & 0.10 & 3.0113 & 1.4915 & 6.5290 & 4.8647 & 0.3795 & 0.0058 & 0.9746 & 0.1920 \\
        & 0.30 & 3.1037 & 1.5097 & 6.0193 & 5.1801 & 0.4782 & 0.0066 & 1.1124 & 0.1496 \\
        & 0.50 & 3.0275 & 1.4947 & 5.5969 & 5.2227 & 0.5249 & 0.0071 & 1.1904 & 0.1022 \\
        & 0.60 & 2.8358 & 1.4556 & 5.8074 & 5.5233 & 0.5636 & 0.0084 & 1.0755 & 0.0744 \\
        & 0.70 & 3.3706 & 1.5619 & 5.1903 & 4.8331 & 0.5772 & 0.0071 & 1.5936 & 0.1491 \\
        & 0.80 & 3.3276 & 1.5535 & 4.9495 & 5.2080 & 0.6117 & 0.0079 & 1.5929 & 0.1397 \\
        & 0.90 & 3.7427 & 1.6335 & 4.3507 & 4.8306 & 0.6983 & 0.0069 & 1.9760 & 0.0207 \\
        & 0.95 & 3.8653 & 1.6568 & 4.1340 & 4.7515 & 0.7848 & 0.0069 & 2.0926 & 0.0442 \\
\hline
\rule{0mm}{3mm}$Z=0.006$ & 0.00 & 2.6607 & 1.4191 & 6.2795 & 4.8178 & 0.4218 & 0.0167 & 0.6678 & 0.1142 \\
        & 0.10 & 2.8571 & 1.4600 & 5.7171 & 4.5974 & 0.3637 & 0.0162 & 0.8697 & 0.2075 \\
        & 0.30 & 3.1193 & 1.5128 & 5.2323 & 4.7427 & 0.4892 & 0.0167 & 1.1535 & 0.1483 \\
        & 0.50 & 3.0771 & 1.5045 & 5.2111 & 5.0265 & 0.5610 & 0.0185 & 1.1342 & 0.1446 \\
        & 0.60 & 2.9607 & 1.4815 & 5.1949 & 5.0690 & 0.5576 & 0.0192 & 1.1200 & 0.1409 \\
        & 0.70 & 3.1077 & 1.5105 & 5.2193 & 4.7928 & 0.6010 & 0.0203 & 1.4027 & 0.1191 \\
        & 0.80 & 3.1568 & 1.5202 & 4.9762 & 4.7993 & 0.5864 & 0.0204 & 1.5126 & 0.1071 \\
        & 0.90 & 3.4196 & 1.5714 & 3.9929 & 4.4800 & 0.6119 & 0.0180 & 1.6830 & 0.1459 \\
        & 0.95 & 3.4000 & 1.5676 & 3.8333 & 4.3666 & 0.6419 & 0.0175 & 1.6814 & 0.1237 \\
\hline
\rule{0mm}{3mm}$Z=0.014$ & 0.00 & 2.4065 & 1.3655 & 5.8070 & 4.8193 & 0.2956 & 0.0400 & 0.4983 & 0.2577 \\
        & 0.10 & 2.7714 & 1.4422 & 4.8298 & 4.2007 & 0.3913 & 0.0337 & 0.8906 & 0.1801 \\
        & 0.30 & 3.3153 & 1.5511 & 3.4444 & 3.5356 & 0.5674 & 0.0272 & 1.4047 & 0.1666 \\
        & 0.50 & 3.1867 & 1.5260 & 3.4473 & 3.7215 & 0.6409 & 0.0287 & 1.3235 & 0.1908 \\
        & 0.60 & 3.2256 & 1.5337 & 3.4455 & 3.8259 & 0.5787 & 0.0309 & 1.4612 & 0.1081 \\
        & 0.70 & 3.2558 & 1.5395 & 3.4518 & 3.9198 & 0.5009 & 0.0334 & 1.5509 & 0.0936 \\
        & 0.80 & 2.9022 & 1.4694 & 4.1405 & 4.7817 & 0.5307 & 0.0392 & 1.0723 & 0.1649 \\
        & 0.90 & 2.9977 & 1.4888 & 4.0481 & 4.3689 & 0.6091 & 0.0382 & 1.2925 & 0.1168 \\
        & 0.95 & 3.0261 & 1.4944 & 4.2465 & 4.6714 & 0.5045 & 0.0420 & 1.1951 & 0.1783 \\
\hline
\end{tabular}
\end{center}
\tablefoot{\tablefoottext{1}{The remnant mass is computed following the same method as in \citet{Hirschi2005a} and \citet{Georgy2009a}.}}
\end{table*}%

\section{Conclusions}\label{Conclu}

In this paper, we have presented an extended database of rotating stellar models at three different metallicities, for nine different initial rotation parameters and ten different masses, in the line of the previous large grid of stellar models \citep{Ekstrom2012a}. The computations account accurate following of the angular-momentum content, and for the stellar-wind anisotropy. Moreover, they allow to give an estimate of the amount of mass that the star should lose mechanically in an equatorial disc when rotating at the critical velocity. This database will be particularly useful for constructing synthetic populations of stars, accounting for mass, rotation, and metallicity distributions.

In this first paper, we presented some general results on the behaviour of the models in the HRD, their lifetime, surface velocities, chemical enrichment, and advanced phases behaviour. In the second paper of this series \citep{Granada2012a}, extensive comparison between the results presented here and observation concerning Be stars were done, particularly concerning the mean mass-loss rates during the critical-rotation phase, disc mass, and lifetimes.

\begin{table*}
\caption{Main parameters of A-B stars at $Z=0.014=Z_{\sun}$}
\label{TabListModelsZ014}
\begin{center}
\scalebox{0.5}{
\begin{tabular}{cccc||ccccccc|ccccccc|ccccccc}
\hline\hline
\multicolumn{4}{c||}{\rule{0mm}{3mm}} & \multicolumn{7}{c|}{End of H-burning} & \multicolumn{7}{c|}{End of He-burning} & \multicolumn{7}{c}{End of C-burning}\\
     $M_\text{ini}$ & $\Omega/\Omega_\text{crit, ini}$ & $v_\text{eq}$ & $\bar{v}_\text{MS}$ & $\tau_\text{H}$ & $M$ & $v_\text{eq}$ & $v_\text{eq}/v_\text{crit}$ & $Y_\text{surf}$ & $\text{N}/\text{C}$ & $\text{N}/\text{O}$ & $\tau_\text{He}$ & $M$ & $P_\text{rot}$ & $\Omega_\text{c}/\Omega_\text{s}$ & $Y_\text{surf}$ & $\text{N}/\text{C}$ & $\text{N}/\text{O}$ & $\tau_\text{C}$ & $M$ & $P_\text{rot}$ & $\Omega_\text{c}/\Omega_\text{s}$ & $Y_\text{surf}$ & $\text{N}/\text{C}$ & $\text{N}/\text{O}$ \\
\rule[-1.5mm]{0mm}{3mm}$M_{\sun}$ & & \multicolumn{2}{c||}{km s$^{-1}$} & Myr & $M_{\sun}$ & km s$^{-1}$ & & \multicolumn{3}{c|}{mass fract.} & Myr & $M_{\sun}$ & day & & \multicolumn{3}{c|}{mass fract.} & kyr & $M_{\sun}$ & day & & \multicolumn{3}{c}{mass fract.}\\
\hline
\rule{0mm}{3mm}     $ 15.00$ & $0.00$ & $  0.$ & $  0.$ & $   11.015$ & $ 14.81$ & $  0.$ & $0.000$ & $0.2660$ & $  0.2885$ & $  0.1152$ & $    1.315$ & $ 13.34$ & --                  & --                  & $0.3054$ & $  2.0811$ & $  0.5211$ & $    4.721$ & $ 13.25$ & --                  & --                  & $0.3390$ & $  3.0660$ & $  0.7360$ \\
     $ 15.00$ & $0.10$ & $ 44.$ & $ 31.$ & $   11.661$ & $ 14.79$ & $ 19.$ & $0.043$ & $0.2660$ & $  0.2931$ & $  0.1166$ & $    1.528$ & $ 12.23$ & $2.833\cdot 10^{6}$ & $2.579\cdot 10^{7}$ & $0.3235$ & $  2.3726$ & $  0.6413$ & $    2.868$ & $ 12.14$ & $3.780\cdot 10^{6}$ & $3.192\cdot 10^{9}$ & $0.3480$ & $  3.0057$ & $  0.7913$ \\
\hline
\hline
\end{tabular}}
\end{center}
\tablefoot{The data presented in this Table are: initial mass (column 1), initial rotation rate (column 2), initial equatorial velocity (column 3), mean equatorial velocity during the MS (column 4). For each burning phase, we also give the duration of the phase (columns 5, 12, and 19), the mass at the end of the phase (columns 6, 13, and 20), the equatorial velocity (column 7), or the rotation period at the end of the phase (columns 14, and 21), the ratio of the equatorial velocity to the critical one (column 8) or the of the angular velocity to the critical one at the end of the burning phase (columns 15, and 22), the surface helium mass fraction (columns 9, 16, 23), the ratio N/C (columns 10, 17, and 24) and N/O at the end of the phase (columns 11, 18, and 25).}
\end{table*}

\begin{table*}
\caption{Main parameters of A-B stars at $Z=0.006=Z_\text{LMC}$}
\label{TabListModelsZ006}
\begin{center}
\scalebox{0.5}{
\begin{tabular}{cccc||ccccccc|ccccccc|ccccccc}
\hline\hline
\multicolumn{4}{c||}{\rule{0mm}{3mm}} & \multicolumn{7}{c||}{End of H-burning} & \multicolumn{7}{c||}{End of He-burning} & \multicolumn{7}{c}{End of C-burning}\\
     $M_\text{ini}$ & $\Omega/\Omega_\text{crit, ini}$ & $v_\text{eq}$ & $\bar{v}_\text{MS}$ & $\tau_\text{H}$ & $M$ & $v_\text{eq}$ & $v_\text{eq}/v_\text{crit}$ & $Y_\text{surf}$ & $\text{N}/\text{C}$ & $\text{N}/\text{O}$ & $\tau_\text{He}$ & $M$ & $P_\text{rot}$ & $\Omega_\text{S}/\Omega_\text{C}$ & $Y_\text{surf}$ & $\text{N}/\text{C}$ & $\text{N}/\text{O}$ & $\tau_\text{C}$ & $M$ & $P_\text{rot}$ & $\Omega_\text{S}/\Omega_\text{C}$ & $Y_\text{surf}$ & $\text{N}/\text{C}$ & $\text{N}/\text{O}$ \\
\rule[-1.5mm]{0mm}{3mm}$M_{\sun}$ & & \multicolumn{2}{c||}{km s$^{-1}$} & Myr & $M_{\sun}$ & km s$^{-1}$ & & \multicolumn{3}{c|}{mass fract.} & Myr & $M_{\sun}$ & km s$^{-1}$ & & \multicolumn{3}{c|}{mass fract.} & kyr & $M_{\sun}$ & km s$^{-1}$ & & \multicolumn{3}{c}{mass fract.}\\
\hline
     \rule{0mm}{3mm}$ 15.00$ & $0.00$ & $  0.$ & $  0.$ & $   11.097$ & $ 14.87$ & $  0.$ & $0.000$ & $0.2559$ & $  0.2885$ & $  0.1152$ & $    1.338$ & $ 14.00$ & --                  & --                  & $0.2685$ & $  1.3634$ & $  0.3566$ & $    3.504$ & $ 13.90$ & --                  & --                  & $0.3214$ & $  2.7453$ & $  0.6739$ \\
     $ 15.00$ & $0.10$ & $ 48.$ & $ 33.$ & $   11.786$ & $ 14.85$ & $ 22.$ & $0.049$ & $0.2559$ & $  0.2928$ & $  0.1165$ & $    1.403$ & $ 13.51$ & $5.472\cdot 10^{5}$ & $5.194\cdot 10^{6}$ & $0.2969$ & $  2.2371$ & $  0.5926$ & $    3.122$ & $ 13.41$ & $8.667\cdot 10^{5}$ & $7.305\cdot 10^{8}$ & $0.3318$ & $  3.2222$ & $  0.8165$ \\
\hline
\hline
\end{tabular}}
\end{center}
\tablefoot{Same data as in Tab.~\ref{TabListModelsZ014}}
\end{table*}

\begin{table*}
\caption{Main parameters of A-B stars at $Z=0.002=Z_\text{SMC}$}
\label{TabListModelsZ002}
\begin{center}
\scalebox{0.5}{\begin{tabular}{cccc|ccccccc|ccccccc|ccccccc}
\hline\hline
\multicolumn{4}{c|}{\rule{0mm}{3mm}} & \multicolumn{7}{c|}{End of H-burning} & \multicolumn{7}{c|}{End of He-burning} & \multicolumn{7}{c}{End of C-burning}\\
     $M_\text{ini}$ & $\Omega/\Omega_\text{crit, ini}$ & $v_\text{eq}$ & $\bar{v}_\text{MS}$ & $\tau_\text{H}$ & $M$ & $v_\text{eq}$ & $v_\text{eq}/v_\text{crit}$ & $Y_\text{surf}$ & $\text{N}/\text{C}$ & $\text{N}/\text{O}$ & $\tau_\text{He}$ & $M$ & $P_\text{rot}$ & $\Omega_\text{c}/\Omega_\text{s}$ & $Y_\text{surf}$ & $\text{N}/\text{C}$ & $\text{N}/\text{O}$ & $\tau_\text{C}$ & $M$ & $P_\text{rot}$ & $\Omega_\text{c}/\Omega_\text{s}$ & $Y_\text{surf}$ & $\text{N}/\text{C}$ & $\text{N}/\text{O}$ \\
\rule[-1.5mm]{0mm}{3mm}$M_{\sun}$ & & \multicolumn{2}{c|}{km s$^{-1}$} & Myr & $M_{\sun}$ & km s$^{-1}$ & & \multicolumn{3}{c|}{mass fract.} & Myr & $M_{\sun}$ & day & & \multicolumn{3}{c|}{mass fract.} & kyr & $M_{\sun}$ & day & & \multicolumn{3}{c}{mass fract.}\\
\hline
     \rule{0mm}{3mm}$ 15.00$ & $0.00$ & $  0.$ & $  0.$ & $   10.996$ & $ 14.92$ & $  0.$ & $0.000$ & $0.2509$ & $  0.2885$ & $  0.1152$ & $    1.363$ & $ 14.69$ & --                  & --                  & $0.2538$ & $  1.0804$ & $  0.2851$ & $    2.143$ & $ 14.60$ & --                  & --                  & $0.3078$ & $  2.6821$ & $  0.6375$ \\
     $ 15.00$ & $0.10$ & $ 47.$ & $ 36.$ & $   11.712$ & $ 14.91$ & $ 26.$ & $0.053$ & $0.2509$ & $  0.2955$ & $  0.1174$ & $    1.366$ & $ 14.72$ & $9.711\cdot 10^{4}$ & $9.187\cdot 10^{5}$ & $0.2730$ & $  1.9569$ & $  0.5093$ & $    2.483$ & $ 14.61$ & $1.741\cdot 10^{5}$ & $1.312\cdot 10^{8}$ & $0.3174$ & $  3.4360$ & $  0.8187$ \\
\hline
\hline
\end{tabular}}
\end{center}
\tablefoot{Same data as in Tab.~\ref{TabListModelsZ014}}
\end{table*}

\begin{acknowledgements}
The authors thank the anonymous referee, whose useful comments helped to improve this work. CG acknowledges support from EU-FP7-ERC-2012-St Grant 306901.
\end{acknowledgements}

\bibliographystyle{aa}
\bibliography{MyBiblio}

\end{document}